\documentclass[a4paper,12pt]{article}

\usepackage{graphics, epsfig}

\begin{document}

\author{P. A. Hogan\thanks{E-mail : phogan@ollamh.ucd.ie} and E. M. O'Shea
\thanks{E-mail : emer.oshea.2@student.ucd.ie}\\
\small Mathematical Physics Department\\ \small  National
University of Ireland Dublin, Belfield, Dublin 4, Ireland}

\title{Shear--Free Gravitational Waves in an Anisotropic Universe}
\date{}
\maketitle

\begin{abstract}
We study gravitational waves propagating through an anisotropic
Bianchi I dust--filled universe (containing the
Einstein--de--Sitter universe as a special case). The waves are
modeled as small perturbations of this background cosmological
model and we choose a family of null hypersurfaces in this
space--time to act as the histories of the wave--fronts of the
radiation. We find that the perturbations we generate can describe
pure gravitational radiation if and only if the null hypersurfaces
are shear--free. We calculate the gauge--invariant small
perturbations explicitly in this case. How these differ from the
corresponding perturbations when the background space--time is
isotropic is clearly exhibited.

\end{abstract}
\vskip 2truepc\noindent PACS number(s): 04.30.Nk
\thispagestyle{empty}
\newpage

\section{Introduction}\indent

In a recent paper \cite{ours} shear--free gravitational radiation
analogous to the Bateman waves of electromagnetic theory was
studied propagating through isotropic cosmological models. The
radiation was represented as a perturbation of the cosmological
model. It was analysed using the Ellis--Bruni \cite{EB}
gauge--invariant and covariant perturbation formalism. Explicit
solutions of the linear differential equations for the
gauge--invariant small quantities (the Ellis--Bruni variables)
were derived. These variables divide naturally into so--called
vector quantities and tensor quantities. Examples of the vector
quantities are the spatial gradients of the proper density of
matter, of the isotropic pressure, of the expansion scalar of the
matter world--lines, as well as the vorticity vector associated
with the matter world--lines and the heat flow vector of the
matter distribution. The tensor quantities include the shear of
the matter world--lines and the anisotropic stress of the matter
distribution. For perturbations describing gravitational waves the
important variables needed are the tensor variables \cite{ours},
\cite {HE}. If the cosmological model is isotropic then the shear
of the matter world--lines and the anisotropic stress of the
matter distribution are Ellis--Bruni variables. However if the
cosmological model is \emph{anisotropic} then the matter
world--lines have shear and so the shear tensor is not (in its
entirety at least) available to us as an Ellis--Bruni variable.
This has important implications therefore for the study of
gravitational wave propagation in an anisotropic universe. In this
paper we show how to circumvent this difficulty. In addition we
discover what influence the anisotropy in the universe has on the
gravitational waves.

We are concerned here with deriving perturbations of an
anisotropic cosmological model which can be interpreted as
representing shear--free gravitational radiation. For definiteness
we choose a Bianchi I dust--filled universe (containing the
Einstein--de Sitter universe as a special case) and a simple
family of null hypersurfaces in this space--time to act as the
histories of the wave fronts of the radiation. To begin with these
null hypersurfaces can have shear but we show that the
perturbations we generate can only represent gravitational waves
(with the gauge--invariant part of the perturbed Weyl tensor
having the algebraic structure one associates with pure
gravitational radiation) if the null hypersurfaces are
shear--free. This involves a specialisation of the anisotropic
universe. The perturbed quantities are explicitly calculated in
this case and all are derived from a single complex--valued
function whose dependence on one complex variable is analytic and
arbitrary (a feature always associated with shear--free
gravitational radiation) and whose dependence on two real
variables is determined by a wave equation which can easily be
integrated. If the background cosmological model is isotropic
(Einstein--de Sitter) then the wave equation is \emph{second
order} whereas if the background is anisotropic the wave equation
is \emph{first order}. This difference is directly attributable to
the presence of shear in the matter world--lines in the
anisotropic background.

Density perturbations of anisotropic Bianchi I universes have been
thoroughly examined in the Ellis--Bruni formalism by Dunsby
\cite{Dunsby}. Our study of gravitational wave perturbations of a
Bianchi I universe is complementary to this work and has in part
been inspired by it.

The outline of the paper is as follows: The notation used
throughout this paper and some important basic equations are given
in Appendix A and referred to where appropriate. In Sec. II we
describe the Bianchi I dust--filled universe which will play the
role of a ``background'' space--time for our small perturbations.
In Sec. III we identify the Ellis--Bruni gauge--invariant
variables we shall use and give the equations satisfied by these
variables. Then, in Sec. IV we require these Ellis--Bruni
variables to have an arbitrary dependence on a function (in order
to have the possibility of having gravitational waves with an
arbitrary profile) and insert this dependence into the equations.
The calculation of the perturbed quantities when the null
hypersurfaces, described in the second paragraph of this
introduction, are shear--free is also given in Sec. IV and the
analogous calculations for the shearing hypersurfaces (also
referred to above) are given in Appendix B. Finally, Sec. V is a
discussion in which we summarise and comment on the results
obtained in Sec. IV and in Appendix B.

\setcounter{equation}{0}
\section{An Anisotropic Space--Time}\indent

We are interested in constructing models of gravitational waves
propagating through an anisotropic universe. For clarity we choose
a specific Bianchi type I dust--filled universe which is the
dust--filled counterpart of the Kasner universe (see, for example,
\cite{HT}) . The space--time model has line--element
\begin{equation}\label{2.1}
ds^2=A^2(t)\,dx^2+B^2(t)\,dy^2+C^2(t)\,dz^2-dt^2\ ,
\end{equation}
with $ A(t)=(t-k)^p(t+k)^{\frac{2}{3}-p} $\,, $
B(t)=(t-k)^q(t+k)^{\frac{2}{3}-q} $\,, $
C(t)=(t-k)^r(t+k)^{\frac{2}{3}-r} $\,, where $p,\,q,\,r$ are
constants satisfying $p+q+r=1$ and $p^2+q^2+r^2=1$ and $k$ is a
constant (if $k=0$ the line--element (\ref{2.1}) becomes the
Einstein--de--Sitter line--element).

The world--lines of the dust--particles are the time--like
geodesic integral curves of the vector field
$u^{a}\,\partial/\partial x^a=\partial/\partial t$ (thus $ u^a
=\delta^a_4$ since we shall label the coordinates $ x^1=x $,\,
$x^2=y$,\, $x^3=z$ and $x^4=t$). The proper density is
\begin{equation}\label{2.2}
\mu=\frac{\dot{A}\dot{B}}{AB}\,+\frac{\dot{A}\dot{C}}{AC}\,
+\frac{\dot{B}\dot{C}}{BC}=\frac{4}{3\,(t^2-k^2)}\ ,
\end{equation}
where a dot indicates differentiation with respect to $t$. For
future purposes it is useful to note that the field equations
satisfied by the functions $A,\,B,\,C$ given above are
\begin{equation}\label{2.3}
\frac{\dot{B}\dot{C}}{BC}\,+\frac{\ddot{B}}{B}\,+\frac{\ddot{C}}{C}=0\,,
\end{equation}
\begin{equation}\label{2.4}
\frac{\dot{A}\dot{C}}{AC}\,+\frac{\ddot{A}}{A}\,+\frac{\ddot{C}}{C}=0\,,
\end{equation}\begin{equation}\label{2.5}
\frac{\dot{A}\dot{B}}{AB}\,+\frac{\ddot{A}}{A}\,+\frac{\ddot{B}}{B}=0.
\end{equation}
The geodesic world--lines of the dust particles are twist--free
(vorticity tensor $\omega_{ab}=0$) and have \emph{expansion
scalar}
\begin{equation}\label{2.6}
\theta=\frac{\dot{A}}{A}\,+\frac{\dot{B}}{B}\,+\frac{\dot{C}}{C}=\frac{2t}{t^2-k^{2}}\
,
\end{equation}
and \emph{shear tensor}
\begin{equation}\label{2.7}
\sigma_{ab}=\sigma_{(1)}n_{(1)a}n_{(1)b}\,+\sigma_{(2)}n_{(2)a}n_{(2)b}\,+
\sigma_{(3)}n_{(3)a}n_{(3)b}\ ,
\end{equation}
where $n_{(1)a}=A\,\delta^{1}_{a}$\,,\
$n_{(2)a}=B\,\delta^2_{a}$\,,\ $n_{(3)a}=C\,\delta^3_{a}$ and
\begin{equation}\label{2.8}
\sigma_{(1)}=\frac{\dot{A}}{A}-\frac{1}{3}\,\theta\,,\qquad
\sigma_{(2)}=\frac{\dot{B}}{B}-\frac{1}{3}\,\theta\,,\qquad
\sigma_{(3)}=\frac{\dot{C}}{C}-\frac{1}{3}\,\theta\,,
\end{equation}
with $A,\,B,\,C$ given above and $\theta$ given by (\ref{2.6}).
The vorticity and shear tensors and the expansion scalar are
defined in Appendix A.

We note that $\sigma_{ab}=\sigma_{ba}$ satisfies
$\sigma^{ab}\,u_{b}=0$ and, in addition, $\sigma^{a}{}_{a}=0$
(this is equivalent to $\sigma_{(1)}+\sigma_{(2)}+\sigma_{(3)}=0$
which follows for example, from (\ref{2.6}) and (\ref{2.8})).
Following Ellis \cite{Ellis} it is convenient to define a
\emph{projected covariant derivative} operating on tensors which
are orthogonal to $u^{a}$ on all of their indices. In particular,
if for example $T_{abc}$ is such a tensor, then its projected
covariant derivative, denoted by ${}^{(3)}\nabla_{d}\,T_{abc}$\,,
is defined as
\begin{equation}\label{2.9}
{}^{(3)}\nabla_{d}\,T_{abc}=h^{p}_{d}\,h^{q}_{a}\,h^{r}_{b}\,h^{s}_{c}\,T_{qrs;p}\
,
\end{equation}
where $h^{a}_{b}=\delta^{a}_{b}+u^{a}\,u_{b}$ is the projection
tensor and the semi--colon indicates covariant differentiation.
When this derivative operates on a scalar function $f$ it
naturally specialises to the partial derivative projected
orthogonal to $u^{a}$. Thus
\begin{equation}\label{2.10}
{}^{(3)}\nabla_{a}\,f=h^{b}_{a}\,f_{,b}\ ,
\end{equation}
with a comma denoting the partial derivative. We note that
\begin{equation}\label{2.11}
{}^{(3)}\nabla_{a}\,h_{bc}=0\,,
\end{equation}
and thus the projected covariant derivative commutes with the
operations of lowering and raising indices on tensors orthogonal
to $u^{a}$ using $h_{ab}$ and $h^{ab}$ respectively.

In the exact Bianchi type I universe the projected covariant
derivative of any scalar function vanishes. Thus we have
\begin{equation}\label{2.12}
{}^{(3)}\nabla_{a}\,\theta=0\,,\qquad {}^{(3)}\nabla_{a}\,\mu=0\,.
\end{equation}
Also
\begin{equation}\label{2.13}
{}^{(3)}\nabla_{b}\,\sigma_{(a)}=0\,,
\end{equation}
for $a=1,\,2,\,3$, and hence
\begin{equation}\label{2.14}
{}^{(3)}\nabla_{b}\,\sigma^{2}=0\,,
\end{equation}
where
$\sigma^{2}=\frac{1}{2}\,\sigma_{ab}\,\sigma^{ab}=\sigma_{(1)}^2
+\sigma_{(2)}^2+\sigma_{(3)}^2$. It follows from (\ref{2.7}),
(\ref{2.13}) and the fact that ${}^{(3)}\nabla_{c}\,n_{(\alpha )
}^{a}=0$ with $\alpha =1, 2, 3$ that
\begin{equation}\label{2.15}
{}^{(3)}\nabla_{c}\,\sigma_{ab}=0.
\end{equation}
With $\theta$ and $\sigma_{ab}$ given by (\ref{2.6})and
(\ref{2.7}) respectively, it is easily shown using (\ref{2.3}),
(\ref{2.4}), and (\ref{2.5}) that
\begin{equation}\label{2.16}
\dot{\sigma}_{ab} + \theta\,\sigma_{ab}=0.
\end{equation}
Here and throughout a dot indicates covariant differentiation in
the direction of $u^a$. Thus $\dot\sigma _{ab}=\sigma _{ab;c}u^a$.

With respect to the 4--velocity field defined after (\ref{2.1})
the Weyl tensor with components $C_{abcd}$, is decomposed into an
``electric part'' and a ``magnetic part'' given by
\begin{equation}\label{2.17}
E_{ab}=C_{apbq}\,u^{p}\,u^{q}\,,\qquad
H_{ab}={}^{*}C_{apbq}\,u^{p}\,u^{q}\,.
\end{equation}
Here ${}^{*}C_{apbq}$ is the dual of the Weyl tensor and it is
defined by
${}^{*}C_{apbq}=\frac{1}{2}\,\eta_{ap}{}^{rs}\,C_{rsbq}$, where
$\eta_{abcd}=\sqrt{-g}\,\epsilon_{abcd}$ with
$g={\rm{det}}\,(g_{ab})$ and $\epsilon_{abcd}$ is the Levi--Civita
permutation symbol. The electric part of the Weyl tensor of the
anisotropic universe with line element (\ref{2.1}) is non--zero
and can be written in terms of the shear $\sigma_{ab}$ of the
matter world--lines as
\begin{equation}\label{2.18}
E_{ab} = \frac{2}{3}\,\sigma^{2}\,h_{ab} -
\sigma_{ac}\,\sigma^{c}{}_{b} +
\frac{1}{3}\,\theta\,\sigma_{ab}\,.
\end{equation}
This follows from the shear--propagation equation (see equation
(\ref{3.5}) below) in this space--time using the facts that the
matter world--lines in this case are geodesic, twist--free with
shear satisfying (\ref{2.16}) and the dust matter distribution is
stress--free. The magnetic part of the Weyl tensor vanishes in
this universe and so we have
\begin{equation}\label{2.19}
H_{ab}=0.
\end{equation}

In the rest of this paper we wish to construct perturbations of
this anisotropic universe which we can interpret as gravitational
waves propagating through the universe. The histories of the
wave--fronts will be null hypersurfaces in this anisotropic
(background) space--time. To be specific we shall choose the null
hypersurfaces to have equations
\begin{equation}\label{2.20}
\phi\,(x,t)\equiv\, x-T(t) = \,\rm{constant}\,,
\end{equation}
with $dT/dt=A^{-1}$. It is easy to see that these hypersurfaces
are null with respect to the metric $g_{ab}$ given via the
line--element (\ref{2.1}). Thus
\begin{equation}\label{2.21}
g^{ab}\,\phi_{,a}\,\phi_{,b}=0.
\end{equation}

The integral curves of the vector field $g^{ab}\,\phi_{,b}$ are
the null geodesic generators of the hypersurfaces (\ref{2.20}). In
terms of a convenient null tetrad in the anisotropic background
given by the 1--forms,
\begin{eqnarray}\label{2.22}
k_{a}\,dx^{a}&=&A\,dx-dt\ , \qquad
l_{a}\,dx^{a}=-\frac{1}{2}(A\,dx+dt)\ ,\nonumber\\ m_{a}\,dx^{a} &
= & \frac{1}{\sqrt{2}}\,(B\,dy+i\,C\,dz)\ , \,\,
\bar{m}_{a}\,dx^{a}=\frac{1}{\sqrt{2}}\,(Bdy-iCdz)\ ,
\end{eqnarray}
the  complex shear of the null geodesics tangent to $\phi_{,a}$ is
\begin{equation}\label{2.23}
\sigma\,= \phi _{,a;b}\,m^{a}\,m^{b}\,=
\frac{1}{2A}\,\left(\frac{\dot{B}}{B}-\frac{\dot{C}}{C}\right)\,=
\frac{1}{2A}\,(\sigma_{(2)}-\sigma_{(3)})\ ,
\end{equation}
where we have used (\ref{2.8}) in the last equality. We see that
the null hypersurfaces are \emph{shear--free} if and only if
$B=C$. With the explicit expressions for $A,\,B,\,C$ given
following (\ref{2.1}) we see that this corresponds to putting the
constant $r=q$ and choosing $q$ and $p$ such that $p\,+2\,q=1$ and
$p^2+2\,q^2=1$.

 \setcounter{equation}{0}
\section{Equations for Perturbations}\indent

We wish to study gravitational wave propagation in the anisotropic
universe described in Section II. We use the Ellis--Bruni approach
\cite{EB} which involves working with small gauge--invariant and
covariant quantities which vanish in the background rather than
small perturbations of the background metric. The gauge--invariant
variables used depend on the type of perturbations to be studied
and also on the background space--time. For example, $E_{ab}$
defined by (\ref{2.17}) is an Ellis--Bruni variable for
gravitational wave propagation in isotropic cosmologies
\cite{ours}, but in the present context its background value is
non--zero (it is given by (\ref{2.18})) and hence it is not an
Ellis--Bruni variable for our purposes. However, it is possible to
extract a gauge--invariant part $\tilde{E}_{ab}$ of $E_{ab}$ given
by (see equations (\ref{3.13}) and (\ref{3.14}) below)
\begin{equation}\label{3.1}
\tilde{E}_{ab}:=\frac{1}{2}\,\pi_{ab}-(\dot{\sigma}_{ab}
+\theta\,\sigma_{ab}).
\end{equation}
Here $\pi _{ab}$ is the small anisotropic stress in the perturbed
matter distribution (see Appendix A). For our case the important
Ellis--Bruni variables are $\tilde{E}_{ab}$, $H_{ab}$, $\pi_{ab}$,
$\dot{\sigma}_{ab} +\theta\,\sigma_{ab}$ and
${}^{(3)}\nabla_{c}\,\sigma_{ab}$. We set all other Ellis--Bruni
variables, including $q_{a}$, $\omega_{ab}$,
$h^b_c\,\theta_{,b}$\,, $h^b_c\,\mu_{,b}$, $h^b_c\,p_{,b}$ ($p$ is
the isotropic pressure) and $\dot{u}_a$, equal to zero since we
found in \cite{ours} that it is tensor, not vector, quantities
that describe gravitational wave perturbations (the vector
quantities describe other types of perturbations such as
inhomogeneities in the matter density). We now outline the
equations satisfied by these variables.

When the Ricci identities,
\begin{equation}\label{3.2}
u_{a;dc}-u_{a;cd}=R_{abcd}\,u^b\ ,
\end{equation}
where $R_{abcd}$ is the Riemann curvature tensor, are projected
along $u^a$ and orthogonal to $u^a$ (using the projection tensor)
we obtain \emph{Raychaudhuri's equation},
\begin{equation}\label{3.3}
\dot{\theta}+\frac{1}{3}\,\theta^2-\dot{u}^a{}_{;a}+2\,(\sigma^2-\omega^2)+
\frac{1}{2}(\mu+3\,p)=0\ ,
\end{equation}
(here $\omega^2=\frac{1}{2}\,\omega^{ab}\,\omega_{ab}$) the
\emph{vorticity propagation equation},
\begin{equation}\label{3.4}
h^a_b\,\dot{\omega}^b+\frac{2}{3}\,\theta\,\omega^a=\sigma^a{}_b\,\omega^b+
\frac{1}{2}\,\eta^{abcd}\,u_b\,\dot{u}_{c;d}\ ,
\end{equation}
(where $\omega^b:=\frac{1}{2}\,\eta^{abcd}\,u_b\,\omega_{cd}$ is
the vorticity vector) the \emph{shear propagation equation},
\begin{eqnarray}\label{3.5}
&{}&h^f_a\,h^g_b\,(
\dot{\sigma}_{fg}-\dot{u}_{(f;g)})-\dot{u}_a\,\dot{u}_b+\omega_a\,\omega_b+
\sigma_{af}\,\sigma^f{}_b+\frac{2}{3}\,\theta\,\sigma_{ab}\nonumber\\&{}&
+h_{ab}\,\left (-\frac{1}{3}\,\omega^2-
\frac{2}{3}\,\sigma^2+\frac{1}{3}\,\dot{u}^c{}_{;c}\right )
-\frac{1}{2}\,\pi_{ab} +E_{ab}=0\ ,
\end{eqnarray}
the \emph{(0,$\nu$)-field equation} (this terminology is explained
in \cite{Ellis}),
\begin{equation}\label{3.6}
\frac{2}{3}\,h^a_b\,\theta^{,b}-h^a_b\,\sigma^{bc}{}_{;d}\,h^d_c-
\eta^{acdf}\,u_c\,(\omega_{d;f}+2\,\omega_d\,\dot{u}_f)=q^a\ ,
\end{equation}
the \emph{divergence of vorticity equation},
\begin{equation}\label{3.7}
\omega^a{}_{;b}\,h^b_a=\omega^a\dot{u}_a\ ,
\end{equation}
and the \emph{magnetic part of the Weyl tensor},
\begin{equation}\label{3.8}
H_{ab}=2\,\dot{u}_{(a}\,\omega_{b)}-h^t_a\,h^s_b\,(\omega_{(t}{}^{g;c}+\sigma_{(t}{}^{g;c})
\,\eta_{s)fgc}\,u^f\ .
\end{equation}
Next we consider the conservation equation,
\begin{equation}\label{3.9}
T^{ab}{}_{;\,b}=0\ ,
\end{equation}
where $T^{ab}$ is the energy-momentum-stress tensor (see Appendix
A). This equation projected orthogonal to $u^a$ and along $u^a$
respectively gives the \emph{equations of motion of matter},
\begin{equation}\label{3.10}
(\mu+p)\,\dot{u}^a+h^{ac}\,(p_{,c}+\pi^b{}_{c;b}+\dot{q}_c)+ \left
(\omega^{ab}+\sigma^{ab}+\frac{4}{3}\,\theta\,h^{ab}\right
)\,q_b=0\ ,
\end{equation}
and the \emph{energy conservation} equation,
\begin{equation}\label{3.11}
\dot{\mu}+\theta\,(\mu+p)+\pi_{ab}\,\sigma^{ab}+q^a{}_{;a}+\dot{u}^a\,q_a=0\
.
\end{equation}
Specialising  equations (\ref{3.3})--(\ref{3.8}) and
(\ref{3.10})--(\ref{3.11}) to the problem at hand, with perturbed
quantities listed after (\ref{3.1}) and $q_{a}$, $\omega_{ab}$,
$h^b_c\,\theta_{,b}$\,, $h^b_c\,\mu_{,b}$, $h^b_c\,p_{,b}$ and
$\dot{u}_a$ \emph{all put equal to zero}, we obtain the following
equations: From \emph{Raychaudhuri's equation},
\begin{equation}\label{3.12}
\dot{\theta}+\frac{1}{3}\,\theta^2+2\,\sigma^2+
\frac{1}{2}\,(\mu+3\,p)=0\ .
\end{equation}
From the \emph{shear propagation equation},
\begin{equation}\label{3.13}
E_{ab}=\frac{1}{2}\,\pi_{ab}-\dot{\sigma}_{ab}-\sigma_{af}\,\sigma^f{}_b-
\frac{2}{3}\,\theta\,\sigma_{ab}+\frac{2}{3}\,\sigma^2\,h_{ab}\ .
\end{equation}
With $\tilde{E}_{ab}$ given by (\ref{3.1}) we can write this
equation as
\begin{equation}\label{3.14}
E_{ab}=\tilde{E}_{ab}-\sigma_{af}\,\sigma^f{}_b+\frac{1}{3}\,\theta\,\sigma_{ab}+
\frac{2}{3}\,\sigma^2\,h_{ab}\ .
\end{equation}
We now see explicitly from equations (\ref{2.19}) and (\ref{3.14})
that $\tilde E_{ab}=0$ in the anisotropic background. This
justifies the definition of $\tilde E_{ab}$ in equation
(\ref{3.1}).\vskip 1truepc\noindent
 From the \emph{(0,$\nu$)-field
equation}
\begin{equation}\label{3.15}
{}^{(3)}\nabla_{a}\,\sigma^{ab}=0.
\end{equation}
From the \emph{magnetic part of the Weyl tensor}
\begin{equation}\label{3.16}
H_{ab}=-\,h^s_{(a}\,{}^{(3)}\nabla^c\,\sigma^p{}_{b)}\,\eta_{sfpc}\,u^f\
.
\end{equation}
From the \emph{equations of motion of
matter}
\begin{equation}\label{3.17} h^{ac}\,\pi^b{}_{c;b}=0\ .
\end{equation}
Finally the energy conservation equation gives
\begin{equation}\label{3.18}
\dot{\mu}+\theta\,(\mu+p)+\pi_{ab}\,\sigma^{ab}=0\ .
\end{equation}
The remaining two equations derived from the Ricci identities (the
vorticity propagation equation and the divergence of vorticity
equation) are identically satisfied. We note that equations
(\ref{3.12}) and (\ref{3.18}) are not expressed in terms of the
Ellis--Bruni variables we wish to work with and so are not
immediately useful. However, taking the projected covariant
derivative of these equations and retaining only linear terms
yields: \emph{the projected covariant derivative of Raychaudhuri's
equation},
\begin{equation}\label{3.19}
\sigma_{ab}\,{}^{(3)}\nabla_c\,\sigma^{ab}=0\ ,
\end{equation}
and \emph{the projected covariant derivative of the energy
conservation equation},
\begin{equation}\label{3.20}
\sigma^{ab}\,{}^{(3)}\nabla_{c}\,\pi_{ab}=0\ .
\end{equation}

When the Bianchi identities, written conveniently in the form
\begin{equation}\label{3.21}
C^{abcd}{}_{;d} = R^{c[a;b]}-\frac{1}{6}\,g^{c[a}\,R^{;b]}\ ,
\end{equation}
where $R^{ca}:=R^{cba}{}_b$ are the components of the Ricci tensor
and $R:=R^c{}_c$ is the Ricci scalar, are projected along $u^a$
and orthogonal to $u^a$ we obtain equations for $E_{ab}$ and
$H_{ab}$ which are analogous to Maxwell's equations. The general
forms of these equations are lengthy and to keep this section to a
reasonable length we give them in Appendix A. In terms of the
perturbed quantities given following (\ref{3.1}) and with $q_{a}$,
$\omega_{ab}$, $h^b_c\,\theta_{,b}$, $h^b_c\,\mu_{,b}$,
$h^b_c\,p_{,b}$ and $\dot{u}_a$ all vanishing these equations are:
the \emph{div--E equation} (using (\ref{3.17})),
\begin{equation}\label{3.22}
{}^{(3)}\nabla_a\,\tilde{E}^{ab}=\sigma^{af}\,{}^{(3)}\nabla_a\,\sigma^b{}_f+
\eta^{bapq}\,u_a\,\sigma^d{}_p\,H_{qd}\ ,
\end{equation}
the \emph{div--H equation},
\begin{equation}\label{3.23}
{}^{(3)}\nabla_a\,H^{ab}=-\,\eta^{bapq}\,u_a\,\sigma^d{}_p\,\left(\tilde{E}_{qd}+
\frac{1}{2}\,\pi_{qd}\right)\ ,
\end{equation}
the \emph{$\dot{H}$--equation},
\begin{eqnarray}\label{3.24}
&{}&\dot{H}^{bt}-h^{(b}_a\,\eta^{t)rsd}\,u_r\,\tilde{E}^a{}_{s;d}-\eta^{rsd(t}\,u_r\,{}^{(3)}\nabla_d
\,\sigma^{b)f}\,\sigma_{fs}-\eta^{rsd(t}\,u_r\,\sigma^{b)f}\,{}^{(3)}\nabla_d\,\sigma_{fs}
\nonumber\\ &{}&
-\frac{1}{3}\,\theta\,\left({}^{(3)}\nabla_d\,\sigma^{(b}{}_s\right)\,\eta^{t)rsd}\,u_r-
3\,H^{(t}{}_s\,\sigma^{b)s}+h^{bt}\,H^{dp}\,\sigma_{dp}+\theta\,H^{bt}\nonumber\\&{}&=
-\,\frac{1}{2}\,\eta^{(b}{}_{rad}\,\pi^{t)a;d}\,u^r+\frac{1}{2}\,\eta^{(b}{}_{rad}\,u^{t)}\,u^r
\sigma^{cd}\,\pi^a{}_c\ ,
\end{eqnarray}
and the \emph{$\dot{E}$--equation},
\begin{eqnarray}\label{3.25}
&{}&\left( \frac{1}{3}\,\dot{\theta}-2\,\sigma^2+\frac{1}{2}\,\mu
\right)\,\sigma^{bt}+3\,\sigma^{b}{}_f\,\sigma^f{}_s\,\sigma^{st}-h^{bt}\,\sigma^a{}_f\,\sigma^{fd}\,
\sigma_{da}\nonumber\\&{}&
=-\dot{{\tilde{E}}^{bt}}-\frac{2}{3}\,\sigma_{fg}\,(\dot{\sigma}^{fg}
+\theta\,\sigma^{fg})\,h^{bt}-\theta\,\tilde{E}^{bt}+(\dot{\sigma}^b{}_f+
\theta\,\sigma^b{}_f)\,\sigma^{ft} \nonumber\\ & &
+(\dot{\sigma}^{ft}+\theta\,\sigma^{ft})\,\sigma^b{}_f-\frac{1}{3}\,\theta\,(\dot{\sigma}^{bt}
+\theta\,\sigma^{bt})-h^{(b}_a\,\eta^{t)rsd}\,u_r\,H^a{}_{s;d}
\nonumber\\& &
+3\,\tilde{E}^{(t}{}_c\,\sigma^{b)c}-h^{bt}\,\tilde{E}^{dp}\,\sigma_{dp}+\frac{1}{6}\,h^{bt}
\,\pi^{dp}\,\sigma_{dp}-\frac{1}{2}\,\dot{\pi}^{bt}-\frac{1}{2}\sigma^{c(b}\,\pi^{t)}{}_c
\nonumber\\ & & -\frac{1}{6}\,\theta\,\pi^{bt}\ .
\end{eqnarray}

We note that the right--hand side of equation (\ref{3.25}) is
expressed in terms of Ellis--Bruni variables and thus the left
hand side must be an Ellis--Bruni variable which we denote by
\begin{equation}\label{3.26}
W^{bt}:=\left(\frac{1}{3}\,\dot{\theta}-2\,\sigma^2+\frac{1}{2}\,\mu\right)\,\sigma^{bt}+
3\,\sigma^b{}_f\,\sigma^f{}_s\,\sigma^{st}-h^{bt}\,\sigma^a{}_f\,\sigma^{fd}\,\sigma_{ad}\
.
\end{equation}
Calculating the projected covariant derivative of $W^{bt}$ gives
\begin{eqnarray}\label{3.27}
{}^{(3)}\nabla_{c}\,W^{bt}&=&-3\,\sigma^2\,{}^{(3)}\nabla_c\,\sigma^{bt}+
3\,{}^{(3)}\nabla_c\,(\sigma^b{}_f\,\sigma^f{}_s\,\sigma^{st})\nonumber\\
& & -h^{bt}\,{}^{(3)}\nabla_c
(\sigma^a{}_f\,\sigma^{fd}\,\sigma_{da})\ ,
\end{eqnarray}
where we have used the background values of $\theta$, $\mu$ and
$\sigma^{ab}$ given in the previous section to simplify the
coefficient of ${}^{(3)}\nabla_c\,\sigma^{bt}$. Using (\ref{3.25})
and (\ref{3.1}) we can express $W^{tb}$ in terms of
$\tilde{E}^{bt}$, $H^{bt}$ and $\pi^{bt}$ as
\begin{eqnarray}\label{3.28}
&
&W^{bt}=-\dot{\tilde{E}}^{bt}-\frac{2}{3}\,\theta\,\tilde{E}^{bt}-\frac{1}{2}\,\dot{\pi}^{bt}
-\frac{1}{3}\,\theta\,\pi^{bt}-\frac{1}{6}\,h^{bt}\,\pi^{dp}\,\sigma_{dp}\nonumber\\
& & -\frac{1}{3}\,h^{bt} \,\tilde{E}^{dp}\,\sigma_{dp}
+\frac{1}{2}\,\sigma^{(b}{}_f\,\pi^{t)f}+\tilde{E}^{(t}{}_f\,\sigma^{b)f}-h^{(b}_a\,\eta^{t)rsd}
\,u_r\,H^a{}_{s;d}\ .
\end{eqnarray}
We shall use this expression later in the left--hand side of
(\ref{3.27}).

Next we examine Eq. (\ref{3.23}). Substituting for $H^{ab}$ from
(\ref{3.16}) gives
\begin{eqnarray}\label{3.29}
& &
\eta^{frsd}\,u_r\,\left\{-\frac{1}{2}\,\left({}^{(3)}\nabla_d\,\sigma^b{}_s\right)_{;\,f}+u^b\,
\sigma_{fa}\,{}^{(3)}\nabla_d\,\sigma^a{}_s\right\}
-\frac{1}{2}\,\eta^{brsd}\,\sigma_{ra}\,{}^{(3)}\nabla_d\,\sigma^a{}_s\nonumber\\
& &
-\eta^{brsd}\,u_r\,\left\{\frac{1}{2}\,\left({}^{(3)}\nabla_d\,\sigma^a{}_s\right)_{;\,a}-\sigma^a{}_s\,\left(\tilde{E}_{da}+
\frac{1}{2}\,\pi_{ad}\right)\right\}=0\ .
\end{eqnarray}
Multiplying across by $\eta_{blmq}$ and using
\begin{equation}\label{3.30}
u_b\,\left({}^{(3)}\nabla_d\,\sigma^b{}_s\right)_{;\,l}=-\frac{1}{3}\,\theta\,{}^{(3)}\nabla_d\,
\sigma_{ls}\,-\,\sigma_{bl}\,{}^{(3)}\nabla_d\,\sigma^b{}_s\ ,
\end{equation}
(since ${}^{(3)}\nabla_d\,\sigma^b{}_s$ is orthogonal to $u_b$)
and
\begin{equation}\label{3.31}
\left({}^{(3)}\nabla_m\,\sigma^b{}_q\,\right)_{;\,b}-u_m\,\sigma^{ab}\,{}^{(3)}\nabla_a\,
\sigma_{bq}={}^{(3)}\nabla_b\,\left({}^{(3)}\nabla_m\,\sigma^b{}_q\,\right)\
,
\end{equation}
we find that the $div-H$ equation implies that
\begin{equation}\label{3.32}
{}^{(3)}\nabla_b\,\left({}^{(3)}\nabla_{[m}\,\sigma^b{}_{q]}\right)
+\sigma^a{}_{[m}\,\left(\,\tilde{E}_{q]a}+\frac{1}{2}\,\pi_{q]a}\,\right)=0\
,
\end{equation}
with the square brackets as always denoting skew--symmetrisation.
The converse is also true. This is shown by multiplying
(\ref{3.32}) by $\eta^{rsmq}$ and using
\begin{equation}\label{3.33}
{}^{(3)}\nabla_m\,\sigma^b{}_q-{}^{(3)}\nabla_q\,\sigma^b{}_m=\eta_{qpfm}\,u^p\,H^{bf}\
,
\end{equation}
which is easily derived from (\ref{3.16}). Thus we can replace
(\ref{3.23}) by (\ref{3.32}) without any loss of information.

The projected covariant derivative and the covariant derivative in
the direction of $u^a$ do not commute. The commutation relation
satisfied by these derivatives when acting on $\sigma_{ab}$
follows from the Ricci identities and reads:
\begin{eqnarray}\label{3.34}
& &
{}^{(3)}\nabla_c\,\dot{\sigma}_{ab}-\left({}^{(3)}\nabla_c\,\sigma_{ab}\right)\dot{}=\sigma^s{}_a\,{}^{(3)}
\nabla_b\,\sigma_{cs}-\sigma^s{}_a\,{}^{(3)}
\nabla_s\,\sigma_{cb}+\sigma^s{}_b\,{}^{(3)}
\nabla_a\,\sigma_{cs}\nonumber\\ & &-\sigma^s{}_b\,{}^{(3)}
\nabla_s\,\sigma_{ca}+\sigma^s{}_c\,{}^{(3)}
\nabla_s\,\sigma_{ab}+\frac{1}{3}\,\theta\,{}^{(3)}\nabla_c\,\sigma_{ab}\
.
\end{eqnarray}
In the present context this important equation can be written
solely in terms of gauge--invariant variables. Thus we shall use
it in the form,
\begin{eqnarray}\label{3.35}
& &
{}^{(3)}\nabla_c\,(\,\dot{\sigma}_{ab}+\theta\,\sigma_{ab}\,)=\left({}^{(3)}\nabla_c\,
\sigma_{ab}\right)\dot{}-\sigma^f{}_b\,{}^{(3)}\nabla_f\,\sigma_{ac}-
\sigma^f{}_a\,{}^{(3)}\nabla_f\,\sigma_{bc}\nonumber\\ & &
+\sigma^f{}_b\,{}^{(3)}\nabla_a\,\sigma_{fc}+
\sigma^f{}_a\,{}^{(3)}\nabla_b\,\sigma_{fc}+\sigma^f{}_c\,{}^{(3)}\nabla_f\,\sigma_{ab}+
\frac{4}{3}\,\theta\,{}^{(3)}\nabla_c\,\sigma_{ab}\ .
\end{eqnarray}
Using this equation and (\ref{3.1}) we obtain the following
equation for the projected covariant derivative of
$\tilde{E}_{ab}$ ,
\begin{eqnarray}\label{3.36}
& &
{}^{(3)}\nabla_c\,\tilde{E}_{ab}=\frac{1}{2}\,{}^{(3)}\nabla_c\,\pi_{ab}
-\left({}^{(3)}\nabla_c\,
\sigma_{ab}\right)\dot{}+\sigma^f{}_b\,{}^{(3)}\nabla_f\,\sigma_{ac}+
\sigma^f{}_a\,{}^{(3)}\nabla_f\,\sigma_{bc}\nonumber\\ & &
-\sigma^f{}_b\,{}^{(3)}\nabla_a\,\sigma_{fc}-
\sigma^f{}_a\,{}^{(3)}\nabla_b\,\sigma_{fc}-\sigma^f{}_c\,{}^{(3)}\nabla_f\,\sigma_{ab}-
\frac{4}{3}\,\theta\,{}^{(3)}\nabla_c\,\sigma_{ab}\ .
\end{eqnarray}
Contracting this equation over $a$ and $c$ and making use of
(\ref{3.15}) and (\ref{3.17}) gives
\begin{equation}\label{3.37}
{}^{(3)}\nabla_a\,\tilde{E}^{ab}=0\ .
\end{equation}
With $H_{ab}$ given by (\ref{3.16}) it is easily shown that
(\ref{3.22}) is consistent with this equation. It also follows
from (\ref{3.36}), (\ref{3.19}) and (\ref{3.20}) that
\begin{equation}\label{3.38}
\sigma^{ab}\,{}^{(3)}\nabla_c\,\tilde{E}_{ab}=0\ .
\end{equation}
The equations we will use are (\ref{3.1}), (\ref{3.14}),
(\ref{3.15}), (\ref{3.16}), (\ref{3.17}), (\ref{3.19}),
(\ref{3.20}), (\ref{3.24}), (\ref{3.27}), (\ref{3.28}),
(\ref{3.32}), (\ref{3.35}), (\ref{3.37}) and (\ref{3.38}). These
equations are not independent of each other. For example
(\ref{3.24}) is automatically satisfied and this can be seen as
follows: First using (\ref{3.13}) and (\ref{3.35}) we show that
\begin{eqnarray}\label{3.39}
& &
h^b_a\,\eta^{trsd}\,u_r\,\left(\,E^a{}_s-\frac{1}{2}\,\pi^a{}_s\right)_{;\,d}=\,-\,\eta^{trsd}\,
u_r\,\left({}^{(3)}\nabla_d\,\sigma^b{}_s\right)\dot{}
-\theta\,\eta^{trsd}\,u_r\,{}^{(3)}\nabla_d\,\sigma^b{}_s
\nonumber\\ & &
-\eta^{trsd}\,u_r\,\sigma^f{}_s\,(\,{}^{(3)}\nabla^b\,\sigma_{fd}-{}^{(3)}\nabla_f\,\sigma^b{}_d
+{}^{(3)}\nabla_d\,\sigma^b{}_f-{}^{(3)}\nabla_f\,\sigma^b{}_d) .
\end{eqnarray}
Then substituting for
${}^{(3)}\nabla_d\,\sigma^b{}_f-{}^{(3)}\nabla_f\,\sigma^b{}_d$
from (\ref{3.33}) in this equation we find that
\begin{eqnarray}\label{3.40}
h^{(b}_a\,\eta^{t)rsd}\,u_r\,\left(E^a{}_s-\frac{1}{2}\,\pi^a{}_s\right)_{;\,d}=\dot{H}^{bt}-
3\,\sigma^{(t}{}_d\,H^{b)d}+h^{bt}\,\sigma^{dp}\,H_{dp}-\theta\,H^{bt}.
\nonumber\\
\end{eqnarray}
If we replace $E^a{}_s$ in this equation with the right-hand side
of (\ref{3.14}) the result is Eq. (\ref{3.24}) and thus the
$\dot{H}-equation$ is automatically satisfied. We shall
demonstrate the internal consistencies  of the remaining equations
by finding a solution which satisfies all the equations.

\setcounter{equation}{0}
\section{Shear--Free Gravitational Waves}\indent

We shall now look for solutions ${}^{(3)}\nabla_c\,\sigma_{ab}$,
$\dot{\sigma}_{ab}+\theta\,\sigma_{ab}$ and $\pi_{ab}$ of the
equations given in the previous section for which these variables
depend upon an arbitrary function. We expect that this dependence
of perturbations will describe gravitational waves carrying
arbitrary information. Specifically we assume that
\begin{eqnarray}\label{4.1}
{}^{(3)}\nabla_c\,\sigma_{ab}&=&A_{cab}\,F(\phi )+G_{cab}\,F'(\phi
) \ ,\nonumber\\
\dot{\sigma}_{ab}+\theta\,\sigma_{ab}&=&B_{ab}\,F(\phi )
+C_{ab}\,F'(\phi )\ ,\\\pi_{ab}&=&\Pi_{ab}\,F(\phi )\ ,\nonumber
\end{eqnarray}
where $F$ is an arbitrary real--valued function of its argument
$\phi(x^a)$ and $F'$ denotes the derivative of $F$ with respect to
its argument. This idea of introducing arbitrary functions into
solutions of Einsteins equations describing gravitational waves
goes back to work by Trautman \cite{T}. Its use in the context of
gauge--invariant perturbations of cosmological models was
initiated by Hogan and Ellis \cite{HE} and further developed by
Hogan and O'Shea \cite{ours}. We note that all of the quantities
in (\ref{4.1}) are orthogonal to $u^a$ and $B_{ab}$, $C_{ab}$ and
$\Pi_{ab}$ are trace--free with respect to the background metric
$g_{ab}$ (i.e. $B^a{}_a=0$, $C^a{}_a=0$ and $\Pi^a{}_a=0$).

When we substitute the first two equations in Eq. (\ref{4.1}) into
the commutation relation (\ref{3.35}) we, in effect, obtain
integrability conditions to be satisfied by the right hand sides
of these two equations. Remembering that $F$ is an
\emph{arbitrary} function this substitution results in the
following equations:
\begin{equation}\label{4.2}
C_{ab}\,\lambda_c=G_{cab}\,\dot{\phi}\ ,
\end{equation}
where $\lambda_{c}=h^b_c\,\phi_{,b}$ and $\dot\phi =\phi _{,a}u^a$
,
\begin{eqnarray}\label{4.3}
B_{ab}\,\lambda_c+{}^{(3)}\nabla_c\,C_{ab}
&=&A_{cab}\,\dot{\phi}+\dot{G}_{cab}-\sigma^f{}_b\,G_{fac}
-\sigma^{f}{}_a\,G_{fbc}\nonumber\\ & & +
\sigma^f{}_b\,G_{afc}+\sigma^f{}_a\,G_{bfc}+\sigma^f{}_c\,G_{fab}+\frac{4}{3}\,\theta\,G_{cab}\
,
\end{eqnarray}
and
\begin{eqnarray}\label{4.4}
{}^{(3)}\nabla_{c}\,B_{ab}&=&\dot{A}_{cab}-\sigma^f{}_b\,A_{fac}-\sigma^f{}_a\,A_{fbc}+
\sigma^f{}_b\,A_{afc}\nonumber\\ & &
+\sigma^f{}_a\,A_{bfc}+\sigma^{f}{}_c\,A_{fab}+\frac{4}{3}\,\theta\,A_{cab}\
.
\end{eqnarray}
If we now let
\begin{equation}\label{4.5}
C_{ab}=\dot{\phi}\,s_{ab}\ ,
\end{equation}
 where $s_{ab}\,u^a=0$ and $s^a{}_a =0$ then it immediately follows from (\ref{4.2}) that
\begin{equation}\label{4.6}
G_{cab}=s_{ab}\,\lambda_{c}\ .
\end{equation}
As a result of (\ref{4.5}) and (\ref{4.6}) and also since
\begin{equation}\label{4.7}
{}^{(3)}\nabla_c\,\dot{\phi}=(\phi_{,c})\dot{}+\ddot{\phi}\,u_c+\frac{1}{3}\,\theta\,\lambda_c+
\sigma^p{}_c\,\lambda_p\ ,
\end{equation}
and
\begin{equation}\label{4.8}
(\lambda_c)\dot{}=(\phi_{,c})\dot{}+\ddot{\phi}\,u_c\ ,
\end{equation}
(\ref{4.3}) now reads
\begin{eqnarray}\label{4.9}
B_{ab}\,\lambda_c-A_{cab}\,\dot{\phi}&&=-\dot{\phi}\,{}^{(3)}\nabla_c\,s_{ab}-
\frac{1}{3}\,\theta\,\lambda_c\,s_{ab}+\dot{s}_{ab}\,\lambda_{c}
-\sigma^f{}_b\,\lambda_f\,s_{ac}\nonumber\\ & &
-\sigma^f{}_a\,\lambda_f\,s_{bc}+\sigma^f{}_b\,\lambda_a\,s_{fc}+\sigma^f{}_a\,\lambda_b\,s_{fc}
+\frac{4}{3}\,\theta\,\lambda_{c}\,s_{ab}\ .
\end{eqnarray}
Substituting Eqs. (\ref{4.1}) (with $C_{ab}$ and $G_{cab}$
replaced by (\ref{4.5}) and (\ref{4.6}) respectively) into Eqs.
(\ref{3.1}), (\ref{3.15}), (\ref{3.16}) (\ref{3.17}),
(\ref{3.19}), (\ref{3.20}), (\ref{3.27}), (\ref{3.28}),
(\ref{3.32}), (\ref{3.37}) and (\ref{3.38}) results in the
following list of equations: From the \emph{(0, $\nu$) field
equation},
\begin{equation}\label{4.10}
s^{ab}\,\lambda_b=0\ ,
\end{equation}
\begin{equation}\label{4.11}
A_b{}^{ab}=0\ .
\end{equation}
From the \emph{magnetic part of the Weyl tensor},
\begin{equation}\label{4.12}
H_{ab}=q_{ab}\,F+l_{ab}\,F'\ ,
\end{equation}
where as always $F'={dF}/{d\phi}$ and
\begin{equation}\label{4.13}
q_{ab} = -\,A^{cp}{}_{(\,b}\,\eta_{a)fpc}\,u^f\ ,
\end{equation}
\begin{equation}\label{4.14}
l_{ab}=-\,\lambda^c\,s^p{}_{(\,b}\,\eta_{a)fpc}\,u^f\ .
\end{equation}
From the \emph{equations of motion of matter},
\begin{equation}\label{4.15}
\Pi^{ab}{}_{;b} = u^a\,\pi^{bc}\,\sigma_{bc}\ ,
\end{equation}
\begin{equation}\label{4.16}
\Pi^{ab}\,\phi_{,b} = 0\ .
\end{equation}
From the \emph{projected covariant derivative of Raychaudhuri's
equation},
\begin{equation}\label{4.17}
\sigma^{ab}\,A_{cab} =0\ ,
\end{equation}
\begin{equation}\label{4.18}
\sigma^{ab}\,s_{ab} = 0\ .
\end{equation}
From the \emph{projected covariant derivative of the energy
conservation equation},
\begin{equation}\label{4.19}
\sigma_{ab}\,{}^{(3)}\nabla_c\,\Pi^{ab} = 0\ ,
\end{equation}
\begin{equation}\label{4.20}
\sigma_{ab}\,\Pi^{ab} = 0\ .
\end{equation}
From (\ref{3.1})
\begin{equation}\label{4.21}
\tilde{E}_{ab}=(\frac{1}{2}\,\Pi_{ab}-B_{ab})\,F-\dot{\phi}\,s_{ab}\,F'\
.
\end{equation}
From (\ref{3.32})
\begin{eqnarray}\label{4.22}
& &
{}^{(3)}\nabla_b\,A_m{}^b{}_q-{}^{(3)}\nabla_b\,A_q{}^b{}_m+\sigma^a{}_m\,(\Pi_{qa}-B_{qa})
 \nonumber \\ & & -\sigma^a{}_q(\Pi_{ma}-B_{ma})=0\ ,
\end{eqnarray}
and
\begin{eqnarray}\label{4.23}
& &
A_m{}^b{}_q\,\lambda_b+\lambda_m\,{}^{(3)}\nabla_b\,s^b{}_q+s^b{}_q\,{}^{(3)}\nabla_b\,\lambda_m
-A_q{}^b{}_m\,\lambda_b-\lambda_q{}^{(3)}\nabla_b\,s^b{}_m\nonumber\\
& &
-s^b{}_m\,{}^{(3)}\nabla_b\,\lambda_q-\dot{\phi}\,\sigma^a{}_m\,s_{qa}+\dot{\phi}\,\sigma^a{}_q\,s_{ma}=0\
.
\end{eqnarray}
From (\ref{3.37})
\begin{equation}\label{4.24}
{}^{(3)}\nabla_a\,B_{ab}=0\ ,
\end{equation}
\begin{equation}\label{4.25}
B^{ab}\,\lambda_{a}+\dot{\phi}\,{}^{(3)}\nabla_a\,s^{ab}
+s^{ab}\,{}^{(3)}\nabla_a\,\dot{\phi}=0\ ,
\end{equation}
\begin{equation}\label{4.26}
s^{ab}\,\lambda_{b}=0\ .
\end{equation}
From (\ref{3.38})
\begin{equation}\label{4.27}
\sigma_{ab}\,{}^{(3)}\nabla_c\,B^{ab}=0\ ,
\end{equation}
\begin{equation}\label{4.28}
\sigma_{ab}\,B^{ab} = 0\ .
\end{equation}
From (\ref{3.27})
\begin{eqnarray}\label{4.29}
{}^{(3)}\nabla_c\,W^{bt}&=&
3\,(-\sigma^2\,A_c{}^{bt}+\sigma^b{}_f\,\sigma^f{}_s\,A_c{}^{st}+
\sigma^b{}_f\,\sigma^{st}\,A_c{}^f{}_s+\sigma^f{}_s\sigma^{st}\,A_c{}^b{}_f\nonumber\\
&
&-h^{bt}\,\sigma^d{}_a\,\sigma^{af}\,A_{cfd})\,F+3(-\sigma^2\,s^{bt}+\sigma^b{}_f\,\sigma^f{}_s\,s^{st}
+\sigma^b{}_f\,\sigma^{st}\,s^f{}_s \nonumber\\ & &
+\sigma^f{}_s\,\sigma^{st}\,s^b{}_f-h^{bt}\,\sigma^d{}_a\,\sigma^{af}\,s_{fd})\,\lambda_c\,F'\
.
\end{eqnarray}
Finally, from Eq. (\ref{3.28}) using (\ref{4.12})--(4.14),
(\ref{4.18}), (4.20), (\ref{4.21}) and (4.28) we find that
\begin{eqnarray}\label{4.30}
W^{bt}&=&\{\dot{\Pi}^{bt}-\frac{2}{3}\,\theta\,\Pi^{bt}+\dot{B}^{bt}+\frac{2}{3}\,\theta\,B^{bt}
+\sigma^{(\,b}{}_f\,\Pi^{t\,)f}-B^{(\,t}{}_f\,\sigma^{b\,)f}\nonumber
\\ & & {}^{(3)}\nabla_d\,(A^{dbt}-\frac{1}{2}\,A^{tdb}-\frac{1}{2}\,A^{bdt})
\}F+\{(B^{bt}-\Pi^{bt}+\dot{s}^{bt}\nonumber \\ & &
+\frac{2}{3}\,\theta\,s^{bt}
 -s^{(\,t}{}_f\,\sigma^{b\,)f})\,\dot{\phi}
+\ddot{\phi}\,s^{bt}-A^{dbt}\,\lambda_d-\lambda^d\,{}^{(3)}\nabla_d\,s^{bt}
\nonumber \\ & & -s^{bt}\,{}^{(3)}\nabla_d\,\lambda^d
+\frac{1}{2}\,A^{tdb}\,\lambda_d
+\frac{1}{2}\,\lambda^t\,{}^{(3)}\nabla_d\,s^{db}
+\frac{1}{2}\,s^{db}\,{}^{(3)}\nabla_d\,\lambda^t \nonumber \\ &
&+\frac{1}{2}\,A^{bdt}\,\lambda_d
 +\frac{1}{2}\,s^{dt}\,
{}^{(3)}\nabla_d\,\lambda^b+\frac{1}{2}\,\lambda^b\,{}^{(3)}\nabla_d\,s^{dt}\}F'+\nonumber\\
& & \{\dot{\phi}^2\,s^{bt}-\lambda^d\,\lambda_d\,s^{bt}\}F''\ .
\end{eqnarray}
Replacing $W^{bt}$ in (\ref{4.29}) by the right--hand side of
(\ref{4.30}) and equating the coefficients of $F'''$ yields:
\begin{equation}\label{4.31}
(\dot{\phi}^2\,s^{bt}-\lambda^d\,\lambda_d\,s^{bt})\,\lambda_c=0\
.
\end{equation}
Assuming $\lambda_c\neq0$ this implies that
\begin{equation}\label{4.32}
\phi_{,a}\,\phi^{,a}=0\ .
\end{equation}
Thus the hypersurfaces $\phi\,(x^a)={\rm constant}$ in the
background anisotropic cosmological model are null. The equations
found here by equating the coefficients of $F$, $F'$ and $F''$ are
extremely complicated and we will work with them in a completely
different way below.

For the remainder of this section we seek to construct
perturbations of the anisotropic background cosmological model
which describe \emph{shear--free} gravitational waves having the
null hypersurfaces $\phi (x^a)={\rm constant}$ as the histories of
their wave fronts in the background space--time. For the function
$\phi (x^a)$ in (\ref{4.32}) we shall take $\phi (x, t)$ given by
(\ref{2.20}) and in order to have these null hypersurfaces
shear--free we see from (\ref{2.23}) that we must have $\sigma
_{(2)}=\sigma _{(3)}$. This is a relationship between the
(non--vanishing) principal shears of the background matter
distribution. One immediate consequence of this is that
$\tilde{E}_{ab}$ is no longer the \emph{full} gauge invariant part
of $E_{ab}$. To see this we note that in a general space--time
with metric $g_{ab}$ and a preferred congruence of world--lines
tangent to $u^a$ we can write the shear tensor $\sigma _{ab}$ in
terms of the principal shears $\sigma _{(1)}\ ,\sigma _{(2)}\ ,
\sigma _{(3)}$ as
\begin{equation}\label{4.33}
\sigma_{ab}=\sigma_{(1)}\,n_{(1)\,a}\,n_{(1)\,b}+\sigma_{(2)}\,n_{(2)\,a}\,n_{(2)\,b}
+\sigma_{(3)}\,n_{(3)\,a}\,n_{(3)\,b}\ ,
\end{equation}
where $\sigma_{(1)}+\sigma_{(2)}+\sigma_{(3)}=0$ and
$n_{(1)\,a}$\,, $n_{(2)\,a}$ and $n_{(3)\,a}$ are the unit
orthogonal (space--like) eigenvectors of $\sigma_{ab}$. Then
letting $m^a=(n^a_{(2)}+i\,n^a_{(3)})/\sqrt{2}$ and using
(\ref{4.33}) we can write
\begin{equation}\label{4.34}
\left(\frac{1}{3}\,\theta\,\sigma_{ab}-\sigma_{af}\,\sigma^{f}{}_b-\frac{2}{3}\,\sigma^2\,h_{ab}
\right)\,m^a\,m^b=\left(\sigma_{(1)}+\frac{1}{3}\,\theta\right)\,\sigma_{ab}\,m^a\,m^b\
,
\end{equation}
where
\begin{equation}\label{4.35}
\sigma_{ab}\,m^a\,m^b=\frac{1}{2}\,(\sigma_{(2)}-\sigma_{(3)})\ ,
\end{equation}
is now a gauge--invariant variable which vanishes in the
background space--time in which $\sigma_{(2)}=\sigma_{(3)}$. When
$\sigma_{ab}$ is the perturbed shear $m^a$ is given its background
value which, for our case, is given implicitly in (\ref{2.22}).
Thus we can give $\sigma_{(1)}+\frac{1}{3}\,\theta$ its background
value when (\ref{4.34}) is calculated in first approximation. It
follows from this and (\ref{3.14}) that now in first approximation
the full gauge invariant part of $E_{ab}$ is
\begin{equation}\label{4.36}
E'_{ab}:=\tilde{E}_{ab}+\frac{\dot{A}}{A}\,\sigma_{rs}\,m^r\,m^s\,\bar{m}_a\,\bar{m}_b
+\frac{\dot{A}}{A}\,\sigma_{rs}\,\bar{m}^r\,\bar{m}^s\,m_a\,m_b\ .
\end{equation}

We shall consider pure pure gravity wave perturbations of the
anisotropic background. We do this by requiring that the
\emph{gauge--invariant} perturbed ``magnetic'' and ``electric''
parts ($H_{ab}$ and $E'_{ab}$ respectively) of the Weyl tensor be
type N in the Petrov classification with $\phi^{,a}$ as degenerate
principal null direction. We achieve this provided $E'_{ab}\,\phi
^{,b}=0=H_{ab}\,\phi ^{,b}$. This means that, in light of
(\ref{4.10}) and (\ref{4.16}), we should require
\begin{equation}\label{4.37}
B^{ab}\,\phi_{,b}=0\ ,\qquad l^{ab}\,\phi_{,b}=0\ ,\qquad
q^{ab}\,\phi_{,b}=0\ ,
\end{equation}
with $l^{ab}$\,, $q^{ab}$ given above in (\ref{4.13}) and
(\ref{4.14}). The first of these allows us to write Eq.
(\ref{4.25}) as
\begin{equation}\label{4.38}
{}^{(3)}\nabla_a\,(\dot{\phi}\,s^{ab})=0\ ,
\end{equation}
which we can simplify to
\begin{equation}\label{4.39}
(\dot{\phi}\,s^{ab})_{;b}=0\ .
\end{equation}
We note that it follows from the definition of $l_{ab}$ given in
(\ref{4.14}) that the second of (\ref{4.37}) is identically
satisfied.

Since $s^{ab}$, $\Pi^{ab}$ and $B^{ab}$ are orthogonal to $u^a$
and $\phi^{,a}$ and trace--free with respect to the metric tensor
given via the line--element (\ref{2.1}) each have only two
independent components,
$-s^{22}=s^{33}=\alpha\,(x,\,y,\,z,\,t\,)$,\
$s^{23}=s^{32}=\beta\,(x,\,y,\,z,\,t\,)$,\ $B^{23}=B^{32}$,\
$-B^{22}=B^{33}$,\ $\Pi^{23}=\Pi^{32}$ and $-\Pi^{22}=\Pi^{33}$.
It follows from this and the fact that $\sigma_{(2)}=\sigma_{(3)}$
that (\ref{4.18}), (\ref{4.20}) and (\ref{4.28}) (and hence
(\ref{4.19}) and (\ref{4.27})) are identically satisfied.

Calculation of (\ref{4.39}) shows that $\alpha$, $\beta$ must
satisfy the Cauchy--Riemann equations,
\begin{equation}\label{4.40}
\frac{\partial\,\alpha}{\partial\,y}-\frac{\partial\,\beta}{\partial\,z}=0\
,
\end{equation}
\begin{equation}\label{4.41}
\frac{\partial\,\beta}{\partial\,y}+\frac{\partial\,\alpha}{\partial\,z}=0\
.
\end{equation}
We can write these equations economically as
\begin{equation}\label{4.42}
\frac{\partial}{\partial\,\bar\zeta}(\alpha+i\,\beta)=0\ ,
\end{equation}
with $\zeta=y+i\,z$. In addition, combining  (\ref{4.15}),
(\ref{4.20}), (\ref{4.24}) and (\ref{4.27}) we find that
$\Pi^{ab}$ and $B^{ab}$ must also satisfy the Cauchy--Riemann
equations. Thus we have,
\begin{equation}\label{4.43}
\frac{\partial\,\Pi^{33}}{\partial\,y}-\frac{\partial\,\Pi^{23}}{\partial\,z}=0\
,
\end{equation}
\begin{equation}\label{4.44}
\frac{\partial\,\Pi^{23}}{\partial\,y}+\frac{\partial\,\Pi^{33}}{\partial\,z}=0\
,
\end{equation}
and,
\begin{equation}\label{4.45}
\frac{\partial\,B^{33}}{\partial\,y}-\frac{\partial\,B^{23}}{\partial\,z}=0\
,
\end{equation}\begin{equation}\label{4.46}
\frac{\partial\,B^{23}}{\partial\,y}+\frac{\partial\,B^{33}}{\partial\,z}=0\
.
\end{equation}
This appearance of the Cauchy--Riemann equations is to be expected
when working with shear--free null hypersurfaces \cite {R},
\cite{RT}.

Next we examine the integrability condition (\ref{4.9}). To
evaluate this equation we first need to calculate
${}^{(3)}\nabla_c\,s_{ab}$ and $\dot{s}_{ab}+\theta\,s_{ab}$. The
only non--vanishing Christoffel symbols for the line element
(\ref{2.1}) are $\Gamma^1_{14}=\dot{A}/A$\,,
$\Gamma^2_{24}=\dot{B}/B$\,, $\Gamma^3_{34}=\dot{C}/C$\,,
$\Gamma^4_{11}=A\,\dot{A}$\,, $\Gamma^4_{22}=B\,\dot{B}$ and
$\Gamma^4_{33}=C\,\dot{C}$\,. Using these and the fact that $B=C$
we find that
\begin{equation}\label{4.47}
{}^{(3)}\nabla_\beta\,s_{ab}=0\ , \qquad
{}^{(3)}\nabla_4\,s_{ab}=0\ ,
\end{equation}
where $\beta$ takes values 1, 2, 3 and
\begin{equation}\label{4.48}
\dot{s}_{ab}+\theta\,s_{ab}=\frac{\partial
s_{ab}}{\partial\,t}+\frac{\dot{A}}{A}\,s_{ab}\ .
\end{equation}
Then Eq. (\ref{4.9}) yields the following:
\begin{equation}\label{4.49}
\frac{1}{A}\,A_{1ab}+B_{ab}=\frac{1}{A^2}\,\left(\frac{\partial}{\partial\,x}+\frac{\partial}{\partial\,T}
\right)(A\,s_{ab})\ ,
\end{equation}
with $T(t)$ introduced in (2.20), and
\begin{eqnarray}\label{4.50}
\frac{1}{A}\,A_{2ab}&=&\frac{1}{A}\,\frac{\partial
s_{ab}}{\partial\,y}+(\sigma_{(2)}-\sigma_{(1)})\{\,s_{22}\,(\delta^1_b\,\delta^2_a+\delta^1_a\,
\delta^2_b)\nonumber \\ & &
+s_{23}\,(\delta^1_a\,\delta^3_b+\delta^1_b\, \delta^3_a)\}\ ,
\end{eqnarray}
\begin{eqnarray}\label{4.51}
\frac{1}{A}\,A_{3ab}&=&\frac{1}{A}\,\frac{\partial
s_{ab}}{\partial\,z}+(\sigma_{(2)}-\sigma_{(1)})\{\,s_{23}\,(\delta^1_b\,\delta^2_a+\delta^1_a\,
\delta^2_b)\nonumber
\\ & &+s_{33}\,(\delta^1_a\,\delta^3_b+\delta^1_b\, \delta^3_a)\}\
.
\end{eqnarray}
Since $A_{cab}$ is orthogonal to $u^c$ on all its indices we also
have,
\begin{equation}\label{4.52}
A_{4ab}=0\,, \qquad A_{c4b}=0\,, \qquad A_{ca4}=0\ .
\end{equation}
With these equations for $A_{cab}$ it is easily shown that Eqs.
(\ref{4.11}), (\ref{4.17}) and the last of (\ref{4.37}) are
identically satisfied.

We now turn our attention to (\ref{4.22}) and (\ref{4.23}). We
first note that,
\begin{equation}\label{4.53}
s^b{}_q\,{}^{(3)}\nabla_{b}\,\lambda_m=0\ ,
\end{equation}
and
\begin{equation}\label{4.54}
{}^{(3)}\nabla_a\,s^{ab}=0\ .
\end{equation}
Then (\ref{4.23}) reduces to,
\begin{equation}\label{4.55}
(A_m{}^b{}_q-A_q{}^b{}_m)\,\lambda_b+(\sigma^b{}_q\,s_{mb}-\sigma^b{}_m\,s_{qb})\,\dot{\phi}=0\
.
\end{equation}
For convenience we shall define,
\begin{equation}\label{4.56}
X_{(b)mq}:=A_m{}^b{}_q-A_q{}^b{}_m = -X_{(b)qm}\ .
\end{equation}
This allows us to write (\ref{4.55}) as,
\begin{equation}\label{4.57}
X_{(b)mq}\,\lambda_b+(\sigma^b{}_q\,s_{mb}-\sigma^{b}{}_m\,s_{qb})\,\dot{\phi}=0\
.
\end{equation}
After a simple calculation using (\ref{4.49})--(\ref{4.52}) we
find that the non--zero components of $X_{(b)mq}$ are
$X_{(2)12}=-\,X_{(3)13}$ and $X_{(2)13}=-\,X_{(3)12}$ with
\begin{equation}\label{4.58}
X_{(2)12}=
\frac{1}{B^2}\left\{-A\,B_{22}+\frac{1}{A}\,\left(\frac{\partial}{\partial\,x}+\frac{\partial}
{\partial\,T}\right)(A\,s_{22})-A\,(\sigma_{(2)}-\sigma_{(1)})\,s_{22}\right\}\
,
\end{equation}
and
\begin{equation}\label{4.59}
X_{(2)13}=
\frac{1}{B^2}\left\{-A\,B_{23}+\frac{1}{A}\,\left(\frac{\partial}{\partial\,x}+\frac{\partial}
{\partial\,T}\right)(A\,s_{23})-A\,(\sigma_{(2)}-\sigma_{(1)})\,s_{23}\right\}\
,
\end{equation}
As a result of this $X_{(b)mq}\,\lambda_b=0$ and (\ref{4.57})
reduces to
\begin{equation}\label{4.60}
\sigma^b{}_q\,s_{mb}-\sigma^b{}_m\,s_{qb}=0\ .
\end{equation}
This is clearly an identity since $\sigma_{(2)}= \sigma_{(3)}$ and
$s_{22}= - s_{33}$, $s_{23}=s_{32}$ are the only non--zero
components of $s_{ab}$. Thus we have now shown that (\ref{4.23})
is identically satisfied. That (\ref{4.22}) is also identically
satisfied can be seen as follows: $B^{ab}$ and $\Pi^{ab}$ satisfy
the same equations as $s^{ab}$ and so we have,
\begin{equation}\label{4.61}
\sigma^{a}{}_{m}\,\Pi_{aq}-\sigma^a{}_q\,\Pi_{am}=0\
\end{equation}
\begin{equation}\label{4.62}
\sigma^{a}{}_{m}\,B_{aq}-\sigma^a{}_q\,B_{am}=0\ .
\end{equation}
Then Eq. (\ref{4.22}) becomes
\begin{equation}\label{4.63}
{}^{(3)}\nabla_b\,(A_m{}^b{}_q-A_{q}{}^b{}_m)=0\ ,
\end{equation}
which we write as
\begin{equation}\label{4.64}
X_{(c)mq;c}+\Gamma^d_{mc}\,X_{(c)qd}+\Gamma^c_{dc}\,X_{(d)mq}
+\Gamma^d_{qc}\,X_{(c)dm}=0\ ,
\end{equation}
where $X_{(c)mq}$ is given by (\ref{4.56}). Substituting for
$X_{(c)mq}$ from (\ref{4.58})--(\ref{4.59}) and using the
Christoffel symbols which were listed after (\ref{4.46}) we find
that this equation is an identity.

At this point the equations we have left to satisfy are
(\ref{4.4}), (\ref{4.29}), and (\ref{4.42})--(\ref{4.46}). As we
mentioned previously (\ref{4.29}) is extremely complicated and to
simplify the calculations which follow it is convenient to make
use of the null tetrad in the background space--time which was
introduced in (\ref{2.22}). In terms of this tetrad $s^{ab}$ can
be written (because $s^a{}_a=0$,\, $s^{ab}\,u_b=0$,\,
$s^{ab}\,k_b=0$ and so $s^{ab}\,l_b=0$)
\begin{equation}\label{4.65}
s^{ab}=\bar{s}\,m^a\,m^b+s\,\bar{m}^a\,\bar{m}^b\ ,
\end{equation}
with
\begin{equation}\label{4.66}
\bar{s}=-\,B^2\,(\alpha+i\,\beta)\ .
\end{equation}
Since $\Pi^{ab}$ and $B^{ab}$ also satisfy the conditions given
immediately before (\ref{4.65}) we can  write,
\begin{equation}\label{4.67}
\Pi^{ab}=\bar{\Pi}\,m^a\,m^b+\Pi\,\bar{m}^a\,\bar{m}^b\ ,
\end{equation}
with
\begin{equation}\label{4.68}
\bar{\Pi}=-\,B^2\,(\Pi^{33}+i\,\Pi^{23})\ ,
\end{equation}
and
\begin{equation}\label{4.69}
B^{ab}=\bar{\mathcal{B}}\,m^a\,m^b+\mathcal{B}\,\bar{m}^a\,\bar{m}^b\
,
\end{equation}
where
\begin{equation}\label{4.70}
\bar{\mathcal{B}}=-\,B^2(B^{33}+i\,B^{23})\ .
\end{equation}

We recall that $\sigma_{ab}\,m^a\,m^b$, which was first introduced
in (\ref{4.34}), is a gauge--invariant first order small variable
and we can write it as
\begin{equation}\label{4.71}
\sigma_{ab}\,m^a\,m^b=\kappa\,F(\phi )\ ,
\end{equation}
for some function $\kappa$. Multiplying the second of (\ref{4.1})
by $m^a\,m^b$ and replacing $\sigma_{ab}\,m^a\,m^b$ by the
right--hand side of (\ref{4.71}) we find (after noting that
$\dot{m}^a=0$ since $m^a{}_{;d}=\dot{B}\,B^{-1}\,u^a\,m_d$ and
$m_a$ is orthogonal to $u^a$)that
\begin{equation}\label{4.72}
\kappa=s\ ,
\end{equation}
and
\begin{equation}\label{4.73}
\mathcal{B}=\dot{s}+\theta\,s\ .
\end{equation}
We now have an equation for $\mathcal{B}$ in terms of $s$ but for
clarity we shall refrain from replacing $\mathcal{B}$ in the
equations which follow by the right--hand side of this equation
until the end. Repeating the above calculation on the first of
(\ref{4.1}) yields
\begin{equation}\label{4.74}
A_{cab}\,m^a\,m^b={}^{(3)}\nabla_c\,s\ ,
\end{equation}
which is identically satisfied as a consequence of
(\ref{4.49})--(\ref{4.52}) and (\ref{4.73}).

We also need to express $\tilde{E}^{ab}$ and $H^{ab}$ on this
tetrad. It follows from (\ref{4.21}) and the expressions for
$s^{ab}$, $B^{ab}$ and $\Pi^{ab}$ given above that
\begin{equation}\label{4.75}
\tilde{E}^{ab}=\bar{\tilde{E}}\,m^a\,m^b+\tilde{E}\,\bar{m}^a\,\bar{m}^b\
,
\end{equation}
where
\begin{equation}\label{4.76}
\bar{\tilde{E}}=\left(\frac{1}{2}\,\bar{\Pi}-\bar{\mathcal{B}}\right)\,F+
\left(\frac{1}{A}\,\bar{s}\right)\,F'\ .
\end{equation}
We note in passing that, making use of (\ref{4.71})--(\ref{4.72})
and (\ref{4.74})--(\ref{4.75}), we can write $E'_{ab}$ in
(\ref{4.36}) as
\begin{equation}\label{4.77}
E'_{ab}=\bar{E}'\,m^a\,m^b+E'\,\bar{m}^a\,\bar{m}^b\ ,
\end{equation}
with
\begin{equation}\label{4.78}
\bar{E}'=\left(\frac{1}{2}\,\bar{\Pi}-\bar{\mathcal{B}}+\frac{\dot{A}}{A}\,\bar{s}\right)F
+\frac{1}{A}\,\bar{s}\,F'\ .
\end{equation}
Now we consider $H^{ab}$. First, substituting for $A_{cab}$ from
Eqs. (\ref{4.49})--(\ref{4.52}) in (\ref{4.12}) we observe that
$H^{ab}=0$ except for
\begin{eqnarray}\label{4.79}
H^{22}=-H^{33}&=&\left\{(\sigma_{(2)}-\sigma_{(1)})\,s^{23}+B^{23}-\frac{1}{A^2\,B^4}\,\left(
\frac{\partial}{\partial\,x}+\frac{\partial}{\partial\,T}\right)\,(A\,s_{23})\right\}\,F
\nonumber \\ & & -\frac{1}{A}\,s^{23}\,F'\ ,
\end{eqnarray}
and \begin{eqnarray}\label{4.80}
H^{23}=H^{32}&=&\left\{(\sigma_{(2)}-\sigma_{(1)})\,s^{33}-B^{22}-\frac{1}{A^2\,B^4}\,\left(
\frac{\partial}{\partial\,x}+\frac{\partial}{\partial\,T}\right)\,(A\,s_{33})\right\}\,F
\nonumber \\ & & +\frac{1}{A}\,s^{22}\,F'\ ,
\end{eqnarray}
where $\partial/\partial\,T=A(\partial/\partial\,t)$. Then in
terms of the null tetrad we can write
\begin{equation}\label{4.81}
H^{ab}=\bar{H}\,m^a\,m^b+H\,\bar{m}^a\,\bar{m}^b\ ,
\end{equation}
with
\begin{eqnarray}\label{4.82}
\bar{H}&=&i\,\left\{(\sigma_{(2)}-\sigma_{(1)})\,\bar{s}+\bar{\mathcal{B}}-\frac{1}{A^2\,B^2}\,
\left(\frac{\partial}{\partial\,x}+\frac{\partial}{\partial\,T}\right)\,(A\,B^2\,\bar{s})\right\}
\,F \nonumber \\ & & -i\,\frac{1}{A}\,\bar{s}\,F'\ .
\end{eqnarray}
We note that since $s^{ab}$, $B^{ab}$ and $\Pi^{ab}$ satisfy the
Cauchy--Riemann equations, $\bar{s}$, $\bar{\mathcal{B}}$ and
$\bar{\Pi}$ are independent of $\bar\zeta$. But $\bar H$ and
$\bar{\tilde{E}}$ are derived from these quantities and thus they
are also independent of $\bar\zeta$.

Next we wish to express $W^{bt}$ given by (\ref{3.28}) on the null
tetrad but first we need to examine
$h^{(b}_a\,\eta^{t)rsd}\,u_r\,H^a{}_{s;d}$. Using (\ref{4.81}) and
recalling that $m^a{}_{;\,d}=\dot{B}\,B^{-1}\,u^a\,m_d$ we deduce
that
\begin{equation}\label{4.83}
h^{b}_a\,\eta^{trsd}\,u_r\,H^a{}_{s;d}=\eta^{trsd}\,u_r\,\bar{H}_{,d}\,m^b\,m_s+
\eta^{trsd}\,u_r\,H_{,d}\,\bar{m}^b\,\bar{m}_s\ .
\end{equation}
Taking into account that $\partial\,H/\partial\,\zeta=0$ we find
that, with respect to the null tetrad, we can write
\begin{equation}\label{4.84}
H_{,\,d}=\sqrt{2}\,B^{-1}\,H_{\bar\zeta}\,\bar{m}_d-A^{-1}\,(H_x+H_T)\,u_d+H_x\,\phi_{,\,d}\
,
\end{equation}
where $H_{\bar\zeta}=\partial\,H/\partial\,\bar\zeta$,
$H_x=\partial\,H/\partial\,x$ and $H_T=\partial\,H/\partial\,T$.
Putting this into (\ref{4.83}) yields
\begin{equation}\label{4.85}
h^{b}_a\,\eta^{trsd}\,u_r\,H^a{}_{s;d}=\frac{i}{A}\,
(\bar{H}_x\,m^b\,m^t-H_x\,\bar{m}^b\,\bar{m}^t)\ ,
\end{equation}
which is symmetric in $b$ and $t$. We are now in a position to
write $W^{bt}$ in terms of the null tetrad. We find it convenient
to introduce a new variable $\mathcal{E}$ which is defined by
\begin{equation}\label{4.86}
\tilde{E}^{bt}+\frac{1}{2}\,\Pi^{bt}=\mathcal{E}\,\bar{m}^b\,\bar{m}^t+
\bar{\mathcal{E}}\,m^b\,m^t\ .
\end{equation}
It is easily seen (using (\ref{4.67}) and (\ref{4.76})\,) that
\begin{equation}\label{4.87}
\mathcal{E}=(\Pi-\mathcal{B})\,F+A^{-1}\,s\,F'\ .
\end{equation}
Then replacing $E^{bt}+\frac{1}{2}\Pi^{bt}$ and
$h^{(b}_a\,\eta^{t)rsd}\,u_r\,H^a{}_{s;d}$ in (\ref{3.28}) by
(\ref{4.85}) and (\ref{4.86}) respectively gives (since
${}^{(3)}\nabla_c\,m^a=0$ and $\dot{m}^a=0$)
\begin{eqnarray}\label{4.88}
W^{bt}&=&\left(\,-\dot{\bar{\mathcal{E}}}-\frac{2}{3}\,\theta\,\mathcal{E}-i\,A^{-1}\,\bar{H}_x\right)
\,m^b\,m^t+\bar{\mathcal{E}}\,m_f\,m^{(t}\,\sigma^{b)f} \nonumber
\\
&+&\left(-\dot{\mathcal{E}}-\frac{2}{3}\,\theta\,\mathcal{E}+i\,A^{-1}\,H_x\right)\,
\bar{m}^b\,\bar{m}^t +
\mathcal{E}\,\bar{m}_f\,\bar{m}^{(t}\,\sigma^{b)f}\ .
\end{eqnarray}
But a simple calculation shows that
\begin{equation}\label{4.89}
m_f\,m^{(t}\,\sigma^{b)f}=\sigma_{(2)}\,m^b\,m^t=\left(\frac{\dot{B}}{B}-\frac{1}{3}\,\theta\right)
m^b\,m^t\ ,
\end{equation}
and thus we can write,
\begin{equation}\label{4.90}
W^{bt}=\bar{W}\,m^b\,m^t+W\,\bar{m}^b\,\bar{m}^t\ ,
\end{equation}
where
\begin{equation}\label{4.91}
\bar{W}=-\,\frac{1}{A\,B}\,\frac{\partial}{\partial\,t}(A\,B\,\bar{\mathcal{E}})-\frac{i}{A}\,
\bar{H}_x\ .
\end{equation}
Substituting from (\ref{4.82}) and (\ref{4.87}) reveals,
\begin{equation}\label{4.92}
\bar{W}=P\,F+Q\,F'\ ,
\end{equation}
with
\begin{eqnarray}\label{4.93}
P&=&-\,\frac{1}{AB}\,\frac{\partial}{\partial\,t}\{A\,B\,(-\bar{\mathcal{B}}+\bar{\Pi})\}
+\frac{1}{A}\,(\sigma_{(2)}-\sigma_{(1)})\,\bar{s}_x \nonumber \\
& &
+\frac{1}{A}\,\bar{\mathcal{B}}_x-\frac{1}{A^3B^2}\left(\frac{\partial}{\partial\,x}+
\frac{\partial}{\partial\,T}\right)(A\,B^2\,\bar{s}_x)\ ,
\end{eqnarray}
and
\begin{eqnarray}\label{4.94}
Q &=&
\frac{1}{A}\,(\sigma_{(2)}-\sigma_{(1)})\,\bar{s}-\frac{1}{A^3B^2}\left(\frac{\partial}{\partial\,x}+
\frac{\partial}{\partial\,T}\right)(A\,B^{2}\,\bar{s}) \nonumber
\\ & &
-\frac{1}{A^2}\,\bar{s}_x+\frac{1}{A}\,\bar{\Pi}-\frac{1}{AB}\,\frac{\partial}{\partial\,t}
(B\,\bar{s})\ .
\end{eqnarray}

We now turn our attention to Eq.(\ref{4.29}). We first observe
that since $\sigma_{(2)}=\sigma_{(3)}$,
$\sigma_{(1)}+2\,\sigma_{(2)}=0$, and $s^{22}=-s^{33}$,
$s^{32}=s^{23}$ are the only non--zero components of $s^{ab}$, the
coefficient of $F'$ in Eq.(\ref{4.29}) vanishes identically.
Therefore
\begin{eqnarray}\label{4.95}
{}^{(3)}\nabla_c\,W^{bt}&=&
3\,(-\,\sigma^2\,A_c{}^{bt}+\sigma^b{}_f\,\sigma^f{}_s\,A_c{}^{st}+
\sigma^b{}_f\,\sigma^{st}\,A_c{}^f{}_s \nonumber \\ & &
+\sigma^f{}_s\,\sigma^{st}\,A_c{}^b{}_f-h^{bt}\,\sigma_a{}^d\,\sigma^{af}\,A_{cfd})\,F\
.
\end{eqnarray}
Using (\ref{4.49})--(\ref{4.52}) in this we find, after a lengthy
calculation, that the coefficient of $F$ on the right hand side
vanishes and so we are left with
\begin{equation}\label{4.96}
{}^{(3)}\nabla_c\,W^{bt}=0\ .
\end{equation}
Thus from (\ref{4.90}) (since ${}^{(3)}\nabla _{c}m^a=0$) we have
\begin{equation}\label{4.97} {}^{(3)}\nabla_c\,\bar{W}=0\ ,
\end{equation}
which implies that
\begin{equation}\label{4.98}
\frac{\partial\bar{W}}{\partial x}=0\ ,\qquad
\frac{\partial\bar{W}}{\partial\zeta}=0\ ,\qquad
\frac{\partial\bar{W}}{\partial\bar{\zeta}}=0\ .
\end{equation}
Substitution of $\bar W$ from (\ref{4.92}) into the first of
(\ref{4.98}) and using the arbitrariness of $F(\phi )$ we find
that we must have $P=0$ and $Q=0$. Hence $\bar{W}=0$ and the other
two equations in (\ref{4.98}) are identically satisfied. We find
it convenient now to write,
\begin{equation}\label{4.99}
\bar{s}=-\frac{1}{A}\,\mathcal{G}(\zeta,\,x,\,t)\ ,
\end{equation}
where $\cal{G}$ is an analytic function of $\zeta$. This follows
from (\ref{4.42}) and (\ref{4.66}). Then from (\ref{4.94}) with
$Q=0$ we obtain \emph{an equation for $\bar{\Pi}$}, namely,
\begin{equation}\label{4.100}
\bar{\Pi}=-\frac{2}{A^2}\left(D\,\mathcal{G}+A\,\frac{\dot{B}}{B}\,\mathcal{G}\right)\
.
\end{equation}
Here the operator $D$ is given by
$D=\partial/\partial\,x+A\,\partial/\partial\,t=\partial/\partial\,x+\partial/\partial\,T$.
It follows from this and (\ref{4.68}) that $\Pi^{33}+i\,\Pi^{23}$
is analytic in $\zeta$ and so (\ref{4.43}) and (\ref{4.44}) are
automatically satisfied.

We now consider (\ref{4.93}) with $P=0$. Substituting for
$\bar{\Pi}$ and $\bar{\mathcal{B}}$ from (\ref{4.73}) and
(\ref{4.100}) respectively results in the following
\emph{wave--equation for $\mathcal{G}(\zeta\ , x\ , t)$}:
\begin{equation}\label{4.101}
D^2\,\mathcal{G}+A\left(\frac{\dot{B}}{B}-\frac{\dot{A}}{A}\right)D\,\mathcal{G}=0\
.
\end{equation}
We note that we have made use of the field equations
(\ref{2.3})--(\ref{2.5}) (with $B=C$) to write the equation in
this form.

The only equation remaining to be satisfied is (\ref{4.4}). We
shall first examine the case when $a=1$. Using
(\ref{4.49})--(\ref{4.52}) and recalling the only non--zero
Christoffel symbols for the line--element (\ref{2.1}) with $B=C$
given prior to Eq.(\ref{4.47}) above, we find that (\ref{4.4})
with $a=1$ reduces to
\begin{equation}\label{4.102}
(\sigma_{(2)}-\sigma_{(1)})\left(\frac{\partial}{\partial\,t}(A\,s_{ab})+\frac{\partial\,s_{ab}}
{\partial\,x}-2\,A\frac{\dot{B}}{B}\,s_{ab}\right)=0\ .
\end{equation}
Multiplying this equation by $m^a\,m^b$ and noting that
$m^a{}_{,d}=\dot{B}\,B^{-1}\,m^a\,u_d$ yields
\begin{equation}\label{4.103}
(\sigma_{(2)}-\sigma_{(1)})D\,\mathcal{G}=0\ .
\end{equation}
It is easily checked that Eq.(\ref{4.103}) is equivalent to
Eq.(\ref{4.102}). We are interested here in an anisotropic
universe and for this reason we cannot have
$\sigma_{(2)}=\sigma_{(1)}$ (we already have
$\sigma_{(2)}=\sigma_{(3)}$ and if we also put
$\sigma_{(2)}=\sigma_{(1)}$ the result is an isotropic universe).
Thus (\ref{4.103}) gives us a remarkably simple first order
wave--equation,
\begin{equation}\label{4.104}
D\,\mathcal{G}=0\ ,
\end{equation}
from which we can conclude that $\mathcal{G}=\mathcal{G}(\zeta\ ,
x-T(t))$. Now (\ref{4.101}) is automatically satisfied. We note
that if the background cosmology is isotropic then the
line--element (\ref{2.1}) is simplified to $A=B=C=t^{2/3}=R(t)$
(say). In this case $\sigma _{(2)}=\sigma _{(1)}$ and so
(\ref{4.103}) is satisfied while the wave--equation for
$\mathcal{G}(\zeta\ , x\ , t)$ is (\ref{4.101}) specialised to
\begin{equation}\label{4.105}
D^2\mathcal{G}=0\ ,
\end{equation}
and (\ref{4.100}) becomes
\begin{equation}\label{4.106}
\bar\Pi =-\frac{2}{R^2}\left (D\mathcal{G}+\dot
R\,\mathcal{G}\right )\ .
\end{equation}
These are precisely the equations (5.37) and (5.36) found in
\cite{ours} when our general formulation for studying shear--free
gravitational radiation in isotropic cosmologies is specialised to
apply to perturbations of the Einstein--de Sitter universe.

We have yet to examine (\ref{4.4}) when $a=2,\,3$ or $4$. The case
when $a=4$ is identically satisfied as a consequence of
(\ref{4.52}). To check that (\ref{4.4}) is satisfied when $a=2$ or
$a=3$ we first multiply across by $m^a\,m^b$. Noting that
${}^{(3)}\nabla_c\,m^a=0$ this yields
\begin{eqnarray}\label{4.107}
{}^{(3)}\nabla_c\,\mathcal{B}&=&
\dot{A}_{cab}\,m^a\,m^b-2\,\sigma^f{}_b\,A_{fac}\,m^a\,m^b+2\,\sigma^f{}_b\,A_{afc}\,m^a\,m^b
\nonumber \\ & &
+\sigma^f{}_c\,A_{fab}\,m^a\,m^b+\frac{4}{3}\,\theta\,A_{cab}\,m^a\,m^b\
. \end{eqnarray} with $\mathcal{B}$ given by Eq.(\ref{4.70}). Then
using
\begin{equation}\label{4.108}
\frac{\partial}{\partial\,t}(A_{cab}\,m^a\,m^b)=\left(\frac{\partial\,A_{cab}}{\partial\,t}\right)
m^a\,m^b-2\,\frac{\dot{B}}{B}\,A_{cab}\,m^a\,m^b\ ,
\end{equation}
we find after a straightforward calculation that (\ref{4.107}) is
identically satisfied.

Now that $s^{ab}$, $\Pi^{ab}$ and $B^{ab}$ are known we can
calculate $E'$ and $H$ and form the gauge--invariant electric and
magnetic parts of the perturbed Weyl tensor which are given by
(\ref{4.71}) and (\ref{4.81}) respectively. We can write the
result compactly as
\begin{equation}\label{4.109}
E'_{ab}+i\,H_{ab}=-2\left\{\left(\frac{\dot{A}}{A^2}-\frac{\dot{B}}{AB}\right)\mathcal{G}
+\frac{1}{A^2}\,\frac{\partial}{\partial\,x}(\mathcal{G}\,F)\right\}m_a\,m_b\
.
\end{equation}
The scalar product of this complex--valued tensor with itself
clearly vanishes and this is the algebraic property one expects
the gauge--invariant part of the perturbed Weyl tensor to have if
the perturbations describe gravitational waves [cf. Eq.(5.42) in
\cite{ours} to which Eq.(\ref{4.109}) specialises when the
background cosmology is Einstein--de Sitter].

\setcounter{equation}{0}
\section{Discussion}\indent

The feature that makes this study of shear--free gravitational
radiation propagating through a Bianchi I anisotropic universe so
much more complicated than the corresponding study when the
universe is isotropic \cite{ours} is the unavailability to us of
the full perturbed shear of the matter world--lines as an
Ellis--Bruni variable. This is in turn due to the central role
played by the perturbed shear of the matter world--lines along
with the perturbed anisotropic stress of the matter distribution
in the study of gravitational waves. Nevertheless this paper
demonstrates that these difficulties can be circumvented and the
final result, which is summarised in the perturbed anisotropic
stress (\ref{4.100}) and the gauge--invariant perturbation of the
Weyl tensor (\ref{4.107}), with the function $\mathcal{G}$
involved in both satisfying (\ref{4.101}) and (\ref{4.102}), is
easily surveyable. These perturbed quantities have been calculated
with the aid of the basic Ellis--Bruni variables listed in
(\ref{4.1}). This is because $A_{cab}\,,\ G_{cab}\,,\ B_{ab}\,,\
C_{ab}$ and $\Pi_{ab}$ appearing in (\ref{4.1}) are now all known
in terms of $\mathcal{G}$: with $s_{ab}$ given by (\ref{4.65}) and
(\ref{4.99}) we have $C_{ab}$ and $G_{cab}$ given by (\ref{4.5})
and (\ref{4.6}). In addition $B_{ab}$ is given by (\ref{4.69}),
(\ref{4.73}) and (\ref{4.99}) while $A_{cab}$ is given by
(\ref{4.50})--(\ref{4.52}) and $\Pi_{ab}$ is obtained from
(\ref{4.67}) and (\ref{4.100}). At this level the most significant
difference between the perturbations of the anisotropic universe
compared to those of the isotropic universe is that in the former
case the final wave equation to be satisfied by $\mathcal{G}$ is
\emph{first order} whereas it is \emph{second order} in the latter
case. It is clear from (\ref{4.103}) that the anisotropy in the
shear of the matter world--lines of the background is directly
responsible for this.

Finally we see from (\ref{2.23}) that the null hypersurfaces
$\phi=\rm{constant}$ in the anisotropic background have null
geodesic generators \emph{with shear} provided
$\sigma_{(2)}\neq\sigma_{(3)}$. It is informative to carry out an
analysis in parallel with that of section IV in this case. The
calculations are briefly outlined in Appendix B. The significant
conclusion is that the gauge--invariant part of the perturbed Weyl
tensor does \emph{not} possess all of the algebraic properties
associated with gravitational waves having propagation direction
$\phi_{,\,a}$ in the anisotropic background. For example, we
cannot have $H^{ab}\,\phi_{,\,b}$ vanishing if the null
hypersurfaces $\phi(x^a)=\rm{constant}$ we are working with have
shear. Therefore we cannot interpret the perturbations as
describing gravitational waves in this case. This observation
suggests to us the possibility of constructing a
\emph{non--vacuum} generalisation of the Goldberg--Kerr theorem
\cite{CH} which will also generalise the theorem of Szekeres
\cite{Sz} expressing the non--existence of Petrov Type N
dust--filled universes. This will be discussed on another
occasion.

\noindent
\section*{Acknowledgment}\noindent
One of us (E.O'S) wishes to thank Enterprise Ireland and IRCSET
for financial support.

\vskip 8truepc

\appendix
\section{Notation and Useful Equations}
\setcounter{equation}{0} Throughout this paper we use the notation
and sign conventions of {\cite{Ellis}} and we take the speed of
light to be unity. We consider a four dimensional space--time
manifold with metric tensor components $g_{ab}$\,, in a local
coordinate system $\{x^a\}$, and a preferred congruence of
world--lines tangent to the unit time--like vector field with
components $u^a$ (with $u^a\,u_a\,=-1$). With respect to this
4--velocity field the energy--momentum--stress tensor $T^{ab}$ can
be decomposed as
\begin{equation}\label{A1}
T^{ab}=\mu\,u^{a}\,u^{b}+p\,h^{ab}+q^{a}\,u^{b}+q^{b}\,u^{a}+
\pi^{ab}\ ,
\end{equation}
where
\begin{equation}\label{A2}
h^{ab}=g^{ab}+u^{a}\,u^{b}\ ,
\end{equation}
is the projection tensor and
\begin{equation}\label{A3}
q^{a}\,u_{a}=0, \qquad \pi^{ab}\,u_{a}=0, \qquad \pi^{a}{}_a=0\ .
\end{equation}
Then $\mu$ is interpreted as the total energy density measured by
an observer with 4--velocity $u^{a}$, $q^{a}$ is the energy flow
(such as heat flow) measured by this observer, $p$ is the
isotropic pressure and $\pi^{a}{}_{b}$ is the trace--free
anisotropic stress (due, for example, to viscosity). After
absorbing the coupling constant into the energy--momentum--stress
tensor Einstein's field equations can be written as
\begin{equation}\label{A4}
R_{ab}-\frac{1}{2}\,g_{ab}R=T_{ab}\ ,
\end{equation}
where $R_{ab}=R_{a}{}^c{}_{bc}$ are the components of the Ricci
tensor and $R$ is the Ricci scalar.

We indicate covariant differentiation with a semicolon, partial
differentiation by a comma and covariant differentiation in the
direction of $u^{a}$ by a dot. As usual square brackets denote
skew--symmetrisation and round brackets denote symmetrisation.
Thus the 4--acceleration of the time--like congruence is
\begin{equation}\label{A5}
\dot{u}^{a}:=u^{a}{}_{;b}\,u^{b}.
\end{equation}
With respect to $u^{a}$ and $h_{ab}$\,, $u_{a;b}$ can be
decomposed into
\begin{equation}\label{A6}
u_{a;b}=\omega_{ab}+\sigma_{ab}+\frac{1}{3}\,\theta\,h_{ab}-
\dot{u}_{a}\,u_{b}\ ,
\end{equation}
where
\begin{equation}\label{A7}
\omega_{ab}:=u_{[a;b]}+\dot{u}_{[a}\,u_{b]}\ ,
\end{equation}
is the vorticity tensor of the congruence,
\begin{equation}\label{A8}
\sigma_{ab}:=u_{(a;b)}+\dot{u}_{(a}\,u_{b)}\ ,
\end{equation}
is the shear tensor of the congruence and
\begin{equation}\label{A9}
\theta:=u^{a}{}_{;a}\ ,
\end{equation}
is the expansion (or contraction) of the congruence.

Finally in this section we give the equations obtained when the
Bianchi identities, written compactly as
\begin{equation}\label{A10}
C^{abcd}{}_{;d}=R^{c[a;b]}-\frac{1}{6}\,g^{c[a}\,R^{;b]}
\end{equation}
are projected along and orthogonal to $u^{a}$. They are the {\it
div--$E$ equation},
\begin{eqnarray}\label{A11}
&&h^b_g\,E^{gd}{}_{;f}\,h^f_d+3\,\omega ^s\,H^b_s-\eta
^{bapq}\,u_a\,\sigma ^d{}_p\,H_{qd}= \frac{1}{3}\,h^b_c\,\mu ^{,c}
\nonumber \\ && +\frac{1}{2}\left\{-\pi ^{bd}{}_{;d}+u^b\,\sigma
_{cd}\pi ^{cd} -3\,\omega ^{bd}q_d +\sigma
^{bd}\,q_d-\frac{2}{3}\,\theta\,q^b+\pi ^{bd}\dot u_d \right\},
\end{eqnarray}
the {\it div--$H$ equation},
\begin{eqnarray}\label{A12}
&{}&h^b_g\,H^{gd}{}_{;f}\,h^f_d-3\,\omega ^s\,E^b_s+\eta
^{bapq}\,u_a\,\sigma ^d{}_p\,E_{qd}= (\mu +p)\,\omega
^b\nonumber\\ &{}&+\frac{1}{2}\,\eta
^{b}{}_{qac}\,u^q\,q^{a;c}+\frac{1}{2}\,\eta
^{b}{}_{qac}\,u^q\,(\omega ^{dc} +\sigma ^{dc})\,\pi ^{a}{}_{d}\ ,
\end{eqnarray}
the {\it $\dot E$--equation},
\begin{eqnarray}\label{A13}
&{}&h^b_f\,\dot E^{fg}\,h^t_g+h^{(b}_a\,\eta
^{t)rsd}\,u_rH^a_{s;d}- 2\,H^{(b}_s\,\eta ^{t)drs}\,u_d\dot
u_r\nonumber\\ &{}&-E^{(t}_s\,\omega ^{b)s} -3\,E^{(t}_s\,\sigma
^{b)s}+h^{tb}\,E^{dp}\,\sigma _{dp}
+\theta\,E^{bt}=-\frac{1}{2}\,(\mu +p)\,\sigma ^{tb}\nonumber\\
&{}&-\frac{1}{6}\,h^{tb}\,\{\dot\mu +\theta\,(\mu +p)\}
-\,q^{(b}\,\dot u^{t)}- \frac{1}{2}\,u^{(b}\,\dot
q^{t)}-\frac{1}{2}\,q^{(t;b)}\nonumber\\ &{}&+\frac{1}{2}\{\omega
^{c(b}+\sigma ^{c(b}\}\,u^{t)}\,q_c+\frac{1}{6}\,\theta\,u^{(t}\,
q^{b)}-\frac{1}{2}\,\dot{\pi}^{bt} +\pi ^{c(b}\,u^{t)}\,\dot
u_c\nonumber\\ &{}&-\frac{1}{2}\,\{\omega ^{c(b}+\sigma
^{c(b}\}\,\pi ^{t)}{}_c-\frac{1}{6}\,\theta\,\pi ^{bt}\ ,
\end{eqnarray}
and the {\it $\dot H$--equation},
\begin{eqnarray}\label{A14}
&{}&h^b_f\,\dot H^{fg}\,h^t_g-h^{(b}_a\,\eta
^{t)rsd}\,u_rE^a_{s;d}+ 2\,E^{(b}_s\,\eta ^{t)drs}\,u_d\dot
u_r\nonumber\\ &{}&-H^{(t}_s\,\omega ^{b)s} -3\,H^{(t}_s\,\sigma
^{b)s}+h^{tb}\,H^{dp}\,\sigma _{dp}
+\theta\,H^{bt}=-q^{(t}\,\omega ^{b)}\nonumber\\
&{}&-\frac{1}{2}\,\eta ^{(t}{}_{rad}\,\{\omega ^{b)d}+\sigma
^{b)d}\}\,u^r\,q^a- \frac{1}{2}\eta ^{(b}{}_{rad}\,\pi
^{t)a;d}\,u_r\nonumber\\ &{}&+\frac{1}{2}\eta
^{(b}{}_{rad}\,u^{t)}\,u^r\,\{\omega ^{cd}+\sigma ^{cd}\}\,\pi
^a{}_c\ .
\end{eqnarray}

\section{Null Hypersurfaces with Shear}
\setcounter{equation}{0} We note from (\ref{2.23}) that the null
hypersurfaces $\phi = \rm{constant}$ have null geodesic generators
\emph{with shear} provided $\sigma_{(2)}\neq\sigma_{(3)}$. We
briefly describe in this appendix the perturbations analogous to
those of Section IV in this case. Equations
(\ref{4.1})--(\ref{4.32}) still apply and we shall again assume
that $B^{ab}\,\phi_{,b}=0$ (the reason for this is given before
(\ref{4.37})). Now recalling that $s^{ab}$ is orthogonal to $u_a$
and trace--free it follows from (\ref{4.10}) and (\ref{4.18}) that
$s^{ab}$ has only one independent component $s^{23}=s^{32}$.
$B^{ab}$ and $\Pi^{ab}$ are also orthogonal to $u_a$ and
trace--free and satisfy the same equations as $s^{ab}$. Thus they
also have only one independent component, $B^{23}=B^{32}$ and
$\Pi^{23}=\Pi^{32}$. On account of this, the Cauchy--Riemann
equations (\ref{4.40})--(\ref{4.46}) now read,
\begin{equation}\label{B1}
\frac{\partial s^{23}}{\partial y}=0\ ,\qquad \frac{\partial
s^{23}}{\partial z}=0\ ,
\end{equation}
\begin{equation}\label{B2}
\frac{\partial \Pi^{23}}{\partial y}=0\ ,\qquad \frac{\partial
\Pi^{23}}{\partial z}=0\ ,
\end{equation}
\begin{equation}\label{B3}
\frac{\partial B^{23}}{\partial y}=0\ ,\qquad \frac{\partial
 B^{23}}{\partial z}=0\ .
\end{equation}

In the present context ($\sigma_{(2)}\neq\sigma_{(3)}$) equations
(\ref{4.49})--(\ref{4.52}) which we derived from (\ref{4.9}) are
modified to:
\begin{eqnarray}\label{B4}
A_{1ab}&=&(\delta^2_a\,\delta^3_b+\delta^3_a\,\delta^2_b)\left\{-A\,B_{23}+\frac{\partial
s_{23}}{\partial x}+\dot{A}\,s_{23}+A\,\frac{\partial
s_{23}}{\partial t}\right\}\ ,
\\
A_{2ab}&=&A\,s_{23}(\sigma_{(3)}-\sigma_{(1)})(\delta^1_a\,\delta^3_b+\delta^3_a\,\delta^1_b)\
,
\\
A_{3ab}&=&A\,s_{23}(\sigma_{(2)}-\sigma_{(1)})(\delta^1_a\,\delta^2_b+\delta^2_a\,\delta^1_b)\
,
\end{eqnarray}
and
\begin{equation}\label{B7}
A_{4ab}=0\ ,\qquad A_{a4b}=0\ ,\qquad A_{ab4}=0\ .
\end{equation}
It follows immediately from these equations that (\ref{4.11}) and
(\ref{4.17}) are identically satisfied. Using these equations in
(\ref{4.12})--(\ref{4.14}) we find that $H_{ab}=0$ except for
\begin{eqnarray}\label{B8}
H_{11}&=&\frac{A^2}{BC}\,s_{23}\left(\frac{\dot{C}}{C}-\frac{\dot{B}}{B}\right)F\
, \\
H_{22}&=&\left\{\left(\frac{\dot{B}}{C}-\frac{\dot{A}B}{AC}\right)s_{23}+\frac{B}{C}\,B_{23}-
\frac{B}{A^2C}D(A\,s_{23})\right\}F-\frac{B}{AC}\,s_{23}\,F',
\nonumber \\
\\
H_{33}&=&\left\{\left(\frac{\dot{A}C}{AB}-\frac{\dot{C}}{B}\right)s_{23}-\frac{C}{B}\,B_{23}
+\frac{C}{A^2B}\,D(A\,s_{23})\right\}F+\frac{C}{AB}\,s_{23}F'\
.\nonumber \\
\end{eqnarray}
Here the operator $D$ is given by $D=\partial/\partial
x+A\,\partial/\partial t$. We note that as a consequence of these
equations we have
\begin{equation}\label{B11}
H^{ab}\,\phi_{,b}=\delta ^1_a\,\frac{s_{23}}{B}\,(\sigma
_{(3)}-\sigma _{(2)})\,F\neq 0\ . \end{equation} Thus one of the
algebraic properties of the gauge--invariant part of the perturbed
Weyl tensor, namely $H^{ab}\phi _{,b}=0$ , which we expect to hold
if the perturbations describe gravitational waves having the null
hypersurfaces $\phi ={\rm constant}$ as the histories of their
wave--fronts, is not possible unless the null hypersurfaces are
shear--free.

We now consider (\ref{3.28}). First, a straightforward but tedious
calculation gives,
\begin{equation}\label{B12}
h_{a(b}\,\eta_{t)}{}^{rsd}\,u_r\,H^a{}_{s;d}=\frac{1}{2}\,\frac{\partial}{\partial
x}\left(\frac{B}{AC}\,H_{33}-\frac{C}{AB}\,H_{22}\right)(\delta^2_b\,\delta^3_t+\delta^3_b\,
\delta^2_t)\ .
\end{equation}
Then substituting for $\tilde{E}_{ab}$ and $H_{ab}$ from
(\ref{4.21}) and (B8)--(B10) respectively in (\ref{3.28}) and
making use of (\ref{B12}) yields
\begin{equation}\label{B13}
W_{bt}=(PF+QF')(\delta^2_b\,\delta^3_t+\delta^3_b\,\delta^2_t)\ ,
\end{equation}
where (after recalling that
$\sigma_{(1)}+\sigma_{(2)}+\sigma_{(3)}=0$)
\begin{eqnarray}\label{B14}
P&=&-\frac{\partial}{\partial
t}(\Pi_{23}-B_{23})+\left\{\frac{1}{2}\left(\frac{\dot{B}}{B}+\frac{\dot{C}}{C}\right)-
\frac{\dot{A}}{A}\right\}(\Pi_{23}-B_{23})+\frac{1}{A}\frac{\partial
B_{23}}{\partial x} \nonumber \\ & &
-\frac{1}{A^3}\,D\left(A\frac{\partial s_{23}}{\partial
x}\right)-\frac{3}{2A}\,\sigma_{(1)}\,\frac{\partial
s_{23}}{\partial x}\ ,
\end{eqnarray}
and
\begin{eqnarray}\label{B15}
Q&=&\frac{1}{A}\,\Pi_{23}-\frac{1}{A}\frac{\partial s_{23}}
{\partial
t}+\frac{1}{2A}\left(\frac{\dot{B}}{B}+\frac{\dot{C}}{C}\right)s_{23}-\frac{1}{A^3}\,D(As_{23})
\nonumber \\ & &
-\frac{3}{2A}\sigma_{(1)}\,s_{23}-\frac{1}{A^2}\frac{\partial
s_{23}}{\partial x}\ .
\end{eqnarray}

Remembering that the only non--zero components of the background
$\sigma_{ab}$ are $\sigma_{(1)}=\sigma^1{}_1$,
$\sigma_{(2)}=\sigma^2{}_2$ and $\sigma_{(3)}=\sigma^3{}_3$, and
substituting for $A_{cab}$ from (B4)--(B7) we find that, in this
case, the right--hand side of (\ref{4.29}) vanishes identically.
Thus we have
\begin{equation}\label{B16}
{}^{(3)}\nabla_c\,W_{bt}=0\ .
\end{equation}
It is easily shown, using the Cauchy--Riemann equations and the
non--zero Christoffel symbols given before (\ref{4.47}) that this
equation is identically satisfied when $c=2,3$ or $4$. However
when $c=1$ it implies,
\begin{equation}\label{B17}
\frac{\partial W_{bt}}{\partial x}=0\ .
\end{equation}
It follows from this and (\ref{B13}) that $P=0$ and $Q=0$. Similar
to the shear--free case we now find it convenient to define
\begin{equation}\label{B18}
s_{23}=- \frac{1}{A}\,\mathcal{G}(x,t)\ ,
\end{equation}
and
\begin{equation}\label{B19}
B_{23}=-\frac{1}{A}\,\mathcal{H}(x,t)\ .
\end{equation}
Then from (\ref{B15}) with $Q=0$ we obtain
\begin{equation}\label{B20}
\Pi_{23}=-\frac{2}{A^2}\,D\,\mathcal{G}+\frac{1}{A}\left(\frac{\dot{B}}{B}+\frac{\dot{C}}{C}\right)
\mathcal{G}\ .
\end{equation}

Finally we examine Eq. (\ref{4.4}) when the null geodesic
generators of the hypersurfaces have shear (i.e. when
$\sigma_{(2)} \neq \sigma_{(3)}$). When $c=2$ or $c=3$ we find,
after a lengthy calculation utilising (B4)--(B7), that
$\mathcal{H}$ in (\ref{B19}) is now given by
\begin{equation}\label{B21}
\mathcal{H}=2\,\mathcal{G}_t+\frac{1}{A}\,\mathcal{G}_x-\left(\frac{\dot{B}}{B}+
\frac{\dot{C}}{C}\right)\,\mathcal{G}\ ,
\end{equation}
where $\mathcal{G}_t=\partial \mathcal{G}/\partial t$ and
$\mathcal{G}_x=\partial \mathcal{G}/\partial x$. We note that this
equation is valid if and only if $\sigma_{(1)} \neq \sigma_{(2)}$
and $\sigma_{(1)} \neq \sigma_{(3)}$. Then substituting for
$A_{cab}$ and $\mathcal{H}$ from (B4)--(B7) and (\ref{B21})
respectively in (\ref{4.4}) with $c=1$ yields
\begin{equation}\label{B22}
\frac{1}{A^2}\,D^2\,\mathcal{G}-\frac{1}{A}\left(\frac{\dot{B}}{B}-\frac{\dot{C}}{C}\right)D\,
\mathcal{G} -
\left(\frac{\dot{A}\dot{B}}{AB}+\frac{\dot{A}\dot{C}}{AC}-3\frac{\dot{B}\dot{C}}{BC}\right)
\mathcal{G}=0\ ,
\end{equation}
where we have used the field equations (\ref{2.3})--(\ref{2.5}) to
simplify the coefficient of $\mathcal{G}$. Eqs. (\ref{B20}) and
(\ref{B21}) allow us to write,
\begin{equation}\label{B23}
\Pi_{23}-B_{23}=-\frac{1}{A^2}\,\mathcal{G}_x\ ,
\end{equation}
and
\begin{equation}\label{B24}
\frac{1}{A}\,\frac{\partial B_{23}}{\partial
x}=-\frac{2}{A^2}\,G_{tx}-\frac{1}{A^3}\,\mathcal{G}_{xx}+\frac{1}{A^2}\left(
\frac{\dot{B}}{B}+\frac{\dot{C}}{C}\right)\mathcal{G}_x\ .
\end{equation}
Using these two equations it is a simple calculation to show that
(B14) with $P=0$ is identically satisfied. We note that
(\ref{4.4}) with $c=4$ gives $0=0$.

The basic Ellis--Bruni variables listed in (\ref{4.1}) are now
determined in terms of $\mathcal{G}$, with $\mathcal{G}(x, t)$
given by (\ref{B22}).The only non--vanishing component of $s_{ab}$
is given by (\ref{B18}) and of $B_{ab}$ by (\ref{B19}) and
(\ref{B21}), with the corresponding component of $\Pi_{ab}$ given
by (\ref{B20}). As in section IV, $C_{ab}$ and $G_{cab}$ are
obtained from (\ref{4.5}) and (\ref{4.6}) while $A_{cab}$ is given
by (\ref{B4})--(\ref{B7}) in this case.

\end{document}